# Dimethylammonium additives alter both vertical and lateral composition in halide perovskite semiconductors

*Sarthak Jariwala[1,2], Rishi Kumar[3], Giles E. Eperon[4,5], Yangwei Shi[1], David Fenning[3], David S. Ginger[1*].*

1. Department of Chemistry, University of Washington, Seattle, WA 98195, USA
2. Department of Materials Science and Engineering, University of Washington, Seattle, WA 98195, USA
3. Department of Nanoengineering, University of California, San Diego, La Jolla, CA 92093, USA
4. National Renewable Energy Laboratory, Golden, CO 80401, USA
5. Swift Solar Inc., San Carlos, CA 94070, USA

*Corresponding author – dginger@uw.edu

**Abstract**

Adding a large A-site cation, such as dimethylammonium (DMA), to the perovskite growth solution has been shown to improve performance and long-term operational stability of halide perovskite solar cells. To better understand the origins of these improvements, we explore the changes in film structure, composition, and optical properties of a formamidinium (FA), Cs, Pb, mixed halide perovskite following addition of DMA to the perovskite growth solution in the ratio $DMA_{0.1}FA_{0.6}Cs_{0.3}Pb(I_{0.8}Br_{0.2})_3$. Using time-of-flight secondary-ion mass spectrometry (TOF-SIMS) we show that DMA is indeed incorporated into the perovskite, with a higher DMA concentration at the surface. Using a combination of PL microscopy and photo-induced force microscopy-based (PiFM) nanoinfrared (nanoIR), we demonstrate that incorporating DMA into the film leads to increased local heterogeneity in the local bandgap, and clustering of the local formamidinium ($CH_5N_2^+$) composition. In addition, using nano-X-ray diffraction, we demonstrate that DMA incorporation also alters the local structural composition by changing the local $d$-spacing distribution and grain size. Our results suggest that compositional variations in the organic cations at the A-site drive the structural heterogeneity observed in case of DMA incorporated films. Our results also suggest that while current-DMA-additive based approaches do have benefits to operational stability and device performance, process optimization to achieve local compositional and structural homogeneity could further boost both of these gains in performance, bringing further gains to solar cells using DMA additives.

Halide perovskite solar cells have demonstrated rapid improvements in power conversion efficiencies (PCE), with the current single junction cell record being 25.5%, the Si/perovskite tandem record at 29.5%, and the all-perovskite 2-terminal tandem at 24.3%.[1] Nevertheless, these efficiencies are still below the theoretical efficiency of (~32%) for a 1.5 eV single junction perovskite cell, or (~43%) for an ideal perovskite tandem.[2] Many of these limitations are due to open circuit voltage losses in working cells,[3–16] especially in wide-bandgap perovskites needed for optimal tandem cells.[2,7,8,12,17–20]

As a result, researchers have tried many approaches to achieve stable, wide-bandgap perovskites, including both tuning the X site[2,14,21] (particularly with Br incorporation), as well as tuning the A-site,[12,17,22–25] in the general $ABX_3$ structure of halide perovskites. Tuning the A-site can indirectly alter the bandgap by changing the octahedral tilt and thus, the electronic orbital overlap.[12,17,22–24] For instance, Stoddard *et al.* reported the use of a large organic cation such as guanidinium (GA) to increase the bandgap.[17] Rajagopal *et al.* reported the use of phenylethylammonium (PEA) to increase the bandgap as well as improve the relative quasi-fermi level splitting w.r.t the bandgap.[12] Recently, Palmstrom *et al.* demonstrated the use of another large organic cation, dimethylammonium (DMA), to increase the bandgap as well as the long-term operational stability of the resulting devices.[24,25]

Although tuning the A-site using large organic cations has demonstrated improvements in the device performance and operation stability,[12,17,24,25] correlated in many cases with changes to macroscopic crystal structure;[25] the impact of A-site tuning with large organic cations on the microscopic chemical and structural properties remains largely unknown, although it would be expected to be significant. Here, by "large organic cations", we mean cations with a large enough ionic radii that they would not form a 3D perovskite structure according to the empirical observations based on established Goldschmidt's tolerance factor (between 0.8-1.0 for 3D perovskites).[23,26,27] Although the macroscopic shift in bandgap with DMA addition has been well explained in previous work,[25] the origins of performance and stability enhancement following DMA addition remain an open question. It is plausible that DMA has a local chemical as well as a local structural effect. Herein, we investigate the effects of using DMA in the mixed cation and mixed halide system initially reported by Palmstrom and Eperon *et al.*,[24,25] $DMA_{10}FA_{60}Cs_{30}Pb(I_{80}Br_{20})_3$, referred to as DMA10 for brevity. We compare the DMA10 system to control systems without DMA, $FA_{80}Cs_{20}Pb(I_{80}Br_{20})_3$ and $FA_{67}Cs_{33}Pb(I_{80}Br_{20})_3$, referred to as Ctrl 80/20 and Ctrl 67/33, respectively. Ctrl 80/20 has a bandgap of 1.65 eV and is the starting composition from where we tune the DMA incorporation to get DMA10.[24] Ctrl 67/33 has the same FA/Cs ratio as the DMA10 composition and serves as additional control to help investigate the impact of DMA on the A-site, ruling out any influence that could be due to increase in Cs in the solution or lower FA/Cs ratio, as in Ctrl 80/20 (see SI for details on sample fabrication at multiple institutions and UV-Vis- Figure S1).

In order to understand the chemical composition of the DMA10 and control films, we perform time-of-flight secondary-ion mass spectrometry (TOF-SIMS). TOF-SIMS has been used to investigate the chemical composition in various different perovskite compositions.[14,28–31] To analyze the molecular fingerprints in the mass spectra, we first identify and label the relevant characteristic peaks from the raw TOF-SIMS spectra. Next, we use a dimensionality reduction method, principal component analysis (PCA), on the labelled TOF-SIMS spectra. PCA has been widely used for TOF-SIMS spectral decomposition[32–35] as well as for other dimensionality reduction problems.[36] Using PCA, we decompose the labelled TOF-SIMS spectra data from 31 different spectral dimensions to 3 principal components that explain over 99% of the observed variance in the spectral data for different samples (Figure S2). Figure 1a plots the first principal component vs the second principal component for DMA10 and control films. We note that all the DMA10 samples cluster together and are well-separated from each of the different control samples clusters. This well-separated clustering indicates that there are distinct chemical differences between the three different sample groups – DMA10, Ctrl 67/33, and Ctrl 80/20 – as one might expect from their different growth stoichiometries. In other words, adding DMA to the precursor solution leads to chemical variations in the resulting film as measured by the resulting mass-spectra. The largest variations between the DMA10 samples and the control samples occur along the first principal component (Figure 1a). The loadings plot (Figure S3) corresponding to the scores of the first principal component shows the highest loadings are for the peaks with mass to charge (m/z) ratios of 45.05, 46.06, and 73.08, which correspond to FA, DMA, and $C_3H_9N_2^+$. The higher loadings indicate that the variations observed along the first principal component are primarily due to the differences in amounts of these mass numbers present in the different samples.

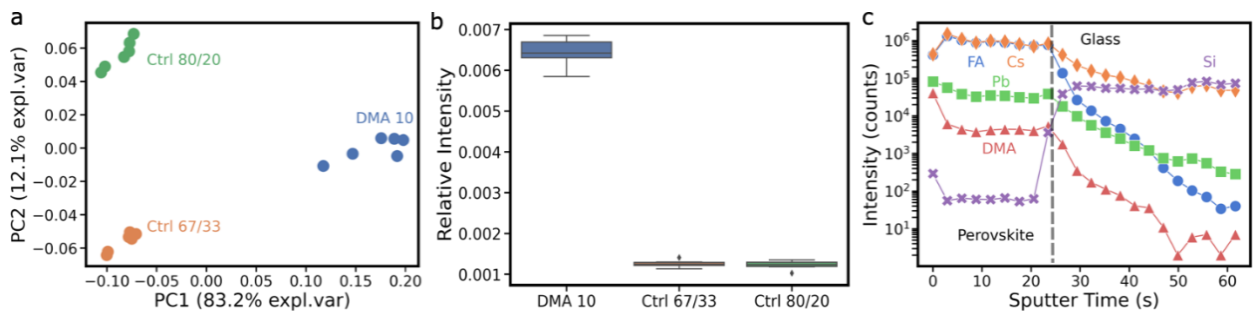

**Figure 1.** TOF-SIMS analysis of halide perovskite films prepared with the compositions $DMA_{10}FA_{60}Cs_{30}Pb(I_{80}Br_{20})_3$ (DMA10), $FA_{67}Cs_{33}Pb(I_{80}Br_{20})_3$ (Ctrl 67/33), and $FA_{80}Cs_{20}Pb(I_{80}Br_{20})_3$ (Ctrl 80/20). a) Principal component analysis (PCA) of TOF-SIMS spectra data collected for DMA 10, Ctrl 67/33 and Ctrl 80/20 thin films. Plot of the first principal component vs the second principal component for the DMA10, Ctrl 67/33, and Ctrl 80/20 samples. b) Intensity of the m/z=46.06 (DMA peak) relative to the total intensity counts from TOF-SIMS spectra for DMA10, Ctrl 67/33, and Ctrl 80/20. c) TOF-SIMS depth profile of the positive ions (FA, Cs, Pb, DMA, Si) in a representative DMA10 film.

Figure 1b shows the normalized relative intensity for the m/z=46.06 peak, which corresponds to DMA. Importantly, we observe that this peak for m/z=46.06 is only present in the samples with DMA added, indicating that DMA is incorporated into the film from the precursor solution despite the larger size (2.72 Å, that would lead to tolerance factor greater than 1). This observation is generally consistent with our previous results.[25] In addition, the TOF-SIMS intensity data for m/z=45.05 peak (FA) (normalized to the total counts) qualitatively agrees with the amount of FA added to the precursor solution (Figure S5) *i.e.,* Ctrl 80/20 has a more intense FA peak than both the Ctrl 67/33 and the DMA10 films (60% FA). The TOF-SIMS data also confirm the same qualitative trends in Cs intensity, with the relative Cs signals determined from TOF-SIMS tracking the concentrations in the precursor solutions (Figure S6).

To understand the chemical variations associated with DMA addition as function of depth in the thin film, we took TOF-SIMS depth profiles. Figure 1c shows a representative depth profile showing the intensity of the FA, Cs, Pb, and DMA ions in DMA-incorporated thin films. Increasing sputter time indicates increasing depth into the film. We observe that the DMA intensity is highest at the surface, and then decreases with increasing depth before reaching a relatively constant level in the bulk of the film. The FA and Cs intensities show the opposite trend from DMA: notably, their concentration is lowest at the surface, and increases as we probe past the surface layer, before reaching a relatively constant value throughout the bulk. We observe similar intensity depth profiles across different samples and regions (Figure S7). We note that despite a higher concentration of DMA at the surface, using XRD we do not observe a secondary 2D phase (Figure S8), consistent with previous reports of no secondary 2D phases at 10% DMA addition.[24] Finally, we also observe relatively constant intensity profiles across the depth for I and Br (Figure S9), indicating there is relatively little variation in halide distribution throughout the depth of the film. The intensity depth profiles for the positive ions along with the TOF-SIMS spectra data suggest that although DMA is incorporated throughout the film, the overall incorporation is non-uniform with a higher concentration of DMA at the top surface. The higher concentration at the top surface suggests that the higher performance for DMA-incorporated films observed by our co-authors[24,25] could possibly originate in part due to surface passivation by DMA.

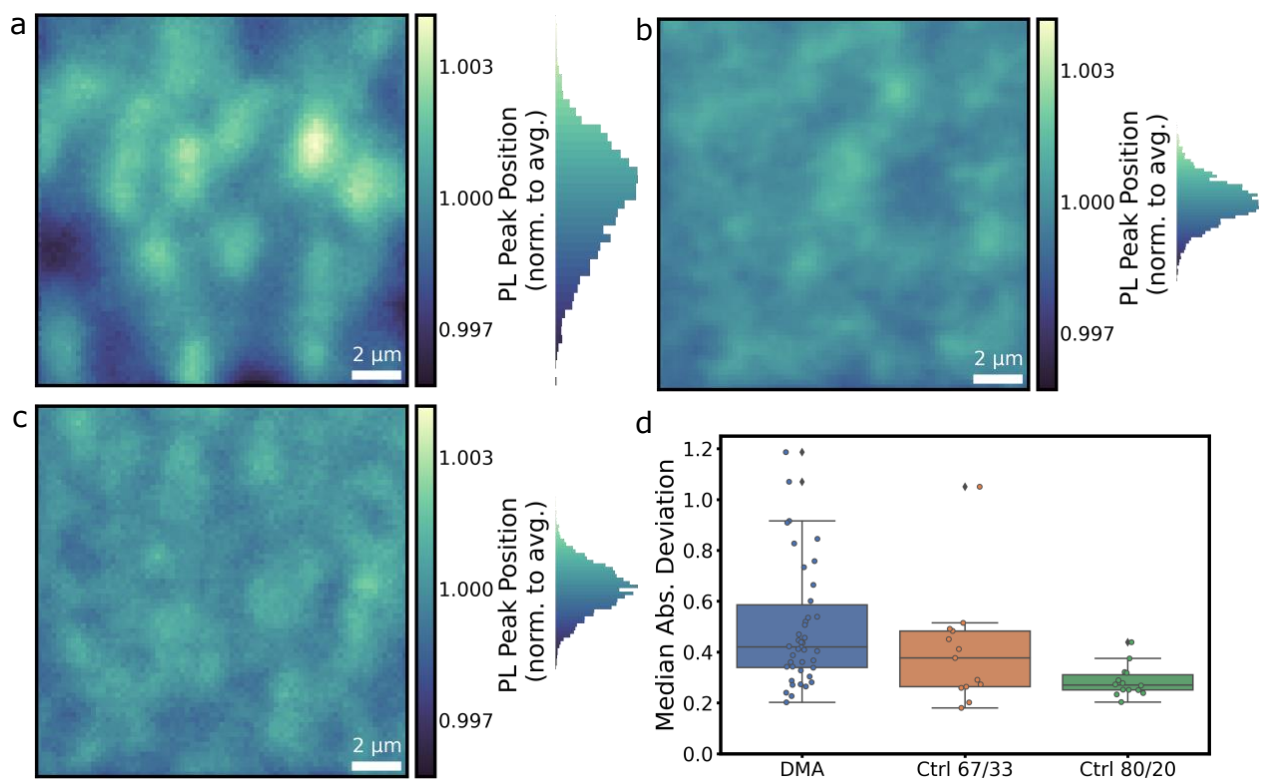

Figure 2. PL spectroscopy images of (a) DMA10, (b) Ctrl 80/20, and (c) Ctrl 67/33 thin films, local PL spectra were acquired at each pixel and normalized to the mean PL peak position for each image. The z-scale in (a), (b) and (c) are kept constant. See Figure S11 for unnormalized images. The histograms in (a), (b), and (c) have the same scale as the z-scale in the respective images. (d) Box plot of median absolute deviation of the local PL spectra for DMA10 (n=42), Ctrl 67/33 (n=13), and Ctrl 80/20 (n=14) thin films. The boxes show quartiles of the data, and the whiskers extend to capture the rest of the distribution. The scatter points are individual data points.

Next, we collect bulk PL spectra for representative DMA and Ctrl 80/20 thin film samples (Figure S10). We observe a broader FWHM for the DMA 10 sample compared to the Ctrl 80/20 sample (Figure S10). This broader PL linewidth following DMA addition is noteworthy because, in general, a narrower PL linewidth should be associated with higher performance.

To investigate origins of the varying PL linewidths, we acquire spatially resolved local PL spectra on DMA10 and control thin films. Figures 2a-c show images of local PL peak position for representative DMA10, Ctrl 80/20, and Ctrl 67/33 thin films, respectively. We find that the DMA incorporated films have a greater variation in the local PL peak position (Figure 2a). The PL spectral maps of the DMA10 films show distinct regions of higher and lower PL peak position, indicative of local compositional variations in the DMA-incorporated film. This local heterogeneity explains the broadening of the PL linewidth observed upon DMA addition both here in this work and by Palmstrom *et al*.[24] In contrast, we do not observe similarly large variations in the PL peak positions for the control films, Ctrl 80/20 (Figure 2b) and Ctrl 67/33 (Figure 2c). Furthermore, we quantify the variations in the local PL peak positions using the *median absolute deviation* metric, which provides the median of the absolute deviation of individual data points from the sample median. Median absolute deviation is more robust against outliers compared to mean and standard deviation.[37] Figure 2d shows the median absolute deviation of the PL peak positions for different DMA incorporated and control thin films. DMA10 films have a 1.5x higher median absolute deviation compared to the control films, quantifying the higher local PL variations in individual DMA incorporated films compared to Ctrl 67/33 and Ctrl 80/20 thin films. This result is consistent with the visual observations made from Figures 2a-c and is strong evidence for the presence of underlying local compositional variations in the DMA10 thin films.

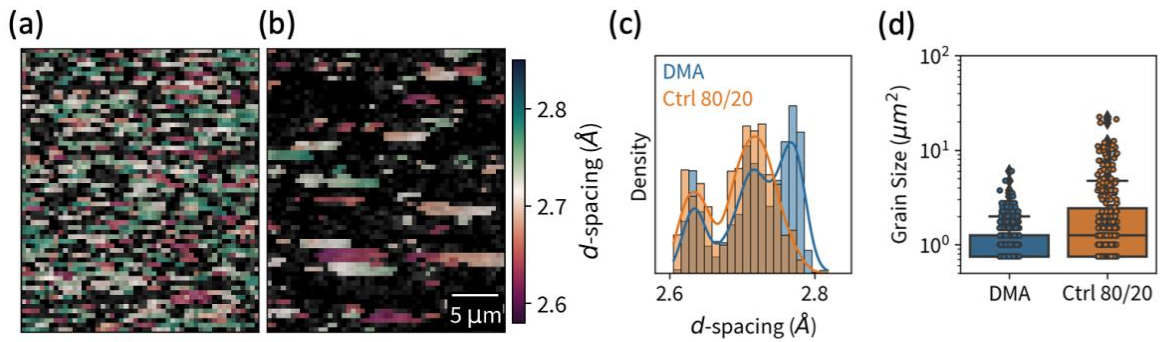

**Figure 3.** Local diffraction intensity overlaid by local *d*-spacing of the (211) reflection for (a) DMA and (b) control films. (c) Distribution of *d*-spacings in (a-b). (d) Distribution of grain sizes including both (211) and (022) orientations. The boxes show quartiles of the data, and the whiskers represent 2.5 quartiles. The scatter points represent individual grain measurements.

To investigate the local structural impact of DMA additive, we characterize the film microstructure by synchrotron nanoprobe X-ray diffraction (nXRD). Using spatially resolved rocking curves taken about the (211) and (022) reflections over 900 µm$^2$ of DMA and Ctrl 80/20 films, we find that the addition of DMA produces nanoscopic domains with higher *d*-spacing than those present in the control sample. The (211) and (022) reflections were specifically chosen for their Bragg angles, which do not overlap with those of any other phases expected in our sample by either phase segregation, decomposition, or unreacted precursors (Figure S12). The nXRD data resolve diffraction intensity across both real and reciprocal space (Figure S13), enabling the separation of signal from individual diffracting grains (Figure S13-S14). Figure 3a-b shows the *d*-spacing extracted from the (211) rocking curves. We note that only regions of the film that are oriented along the <211> direction are visible in Figure 3a-b, and the film is contiguous despite the patchy appearance in the diffraction maps (Figure S17). Figure 3c shows the distribution of

*d*-spacing for DMA10 and Ctrl 80/20 films. We observe that upon DMA incorporation, the (211) d-spacing distribution changes from a bimodal distribution (for Ctrl 80/20 film) to a trimodal distribution, with a third mode at higher *d*-spacing for DMA10 film (Figure 3c). The (022) *d*-spacing distribution also changes from bimodal to trimodal upon DMA incorporation (Figure S16). We also observe that DMA-containing films have reduced grain size, with the Control 80/20 films having a 1.7x median and 3.8x maximum grain size compared to the DMA films across both <022> and <211> orientations (Figure 3d). Here, the grain size is determined by segmentation and measurement of the spatial scattering intensity at each diffracting point in reciprocal space (see SI for details on grain size extraction).

We also acquire nanoscale X-ray fluorescence (nXRF) at the same film location using the same measurement geometry as the nXRD rocking curves (Figure S17). We perform a pixel-by-pixel comparison of local composition to local *d*-spacing (Figure S18), however, we do observe a strong correlation between local halide composition and local *d*-spacing. Analysis of Variance (ANOVA)[38] comparison of composition distributions across film regions belonging to each of the distinct modes in the *d*-spacing distributions at each reflection also fails to show a consistent relationship between local composition and *d*-spacing (Figure S19). The lack of correlation is not surprising given that we are unable to observe DMA or FA using nXRF with our 9 keV probe used. We do, however, note that the local halide composition (I/Pb ratio) and local cesium content (Cs/Pb ratio) are insufficient to explain local *d*-spacing. This observation suggests that compositional variations in the organic cations at the A-site must be driving the variation in *d*-spacing between grains.

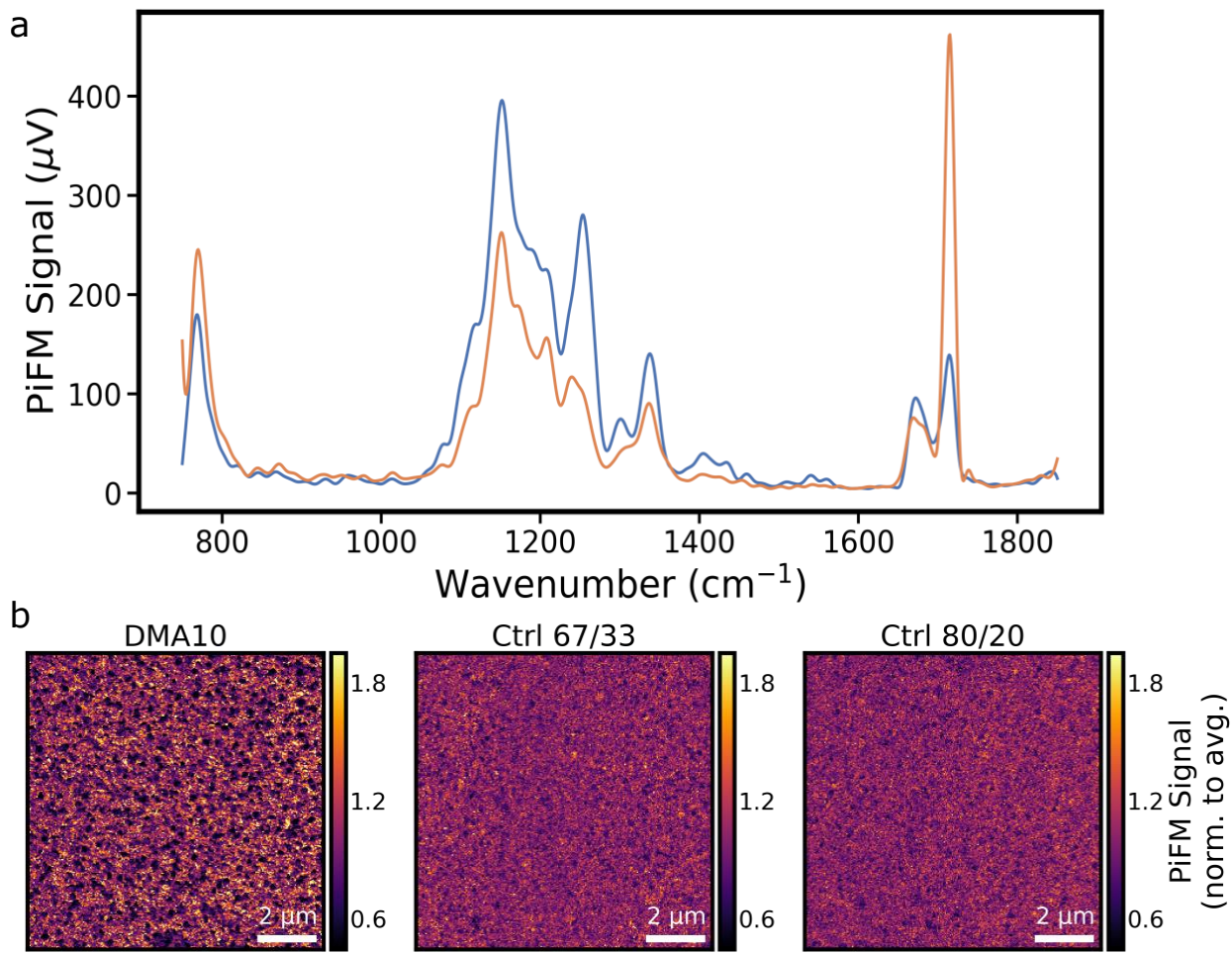

**Figure 4. Photo-induced force microscopy (PiFM) of DMA10 thin films.** (a) PiFM spectra of two regions on a representative DMA10 thin film showing large variations in the peak at 1714 cm$^{-1}$, corresponding to FA. (b) PiFM image of DMA10, Ctrl 67/33 and Ctrl 80/20 films acquired at 1714 cm$^{-1}$, tracking local FA composition. The PiFM signal is normalized to the individual sample average.

In order to explore possible chemical origins for the PL heterogeneity and *d*-spacing changes, we use photo-induced force microscopy (PiFM) to acquire nano-infrared (nanoIR) spectral maps of the films. PiFM uses a tunable IR laser to probe the local surface (~10 nm) chemical vibrations in the material and has been used to study the local composition of organic and inorganic molecules.[36,39–41] Here, we aim to probe the local composition of the organic cations in the DMA10 film. Figure 4a shows the PiFM spectra for two different regions in a representative DMA10 film. From the PiFM spectra, we observe the largest variations in PiFM signal intensity at the 1714 cm$^{-1}$ wavenumber, corresponds to the C=N stretch in FA. We then use the 1714 cm$^{-1}$ FA peak to image the local FA composition in DMA10 and control films. Figure 4b shows the local FA composition for representative DMA10 and control thin films. We observe that the DMA10 films (Figure 4b) have distinct regions of high and low PiFM signal, corresponding to high and low FA concentration, respectively, in the volume probed by the PiFM. The spatial variations in FA composition are greater in DMA10 thin films compared to the Ctrl 80/20 and Ctrl 67/33 thin films (Figure 4b). These observations are consistent across samples from different batches (Figure S21). Figure S22 shows a higher spatial resolution PiFM image at 1714 cm$^{-1}$ for a DMA10 film. We observe that there are entire domains (~300 nm median size) with relatively lower FA signals (~68 $\mu$V on average) than their neighboring regions (~190 $\mu$V on average), indicating a significant level of local FA heterogeneity introduced with DMA addition. We note that these local FA compositional variations observed in the DMA-incorporated samples are *not* simply due to the higher Cs loadings (Figure S23) in these films. We confirm this conclusion by examining Ctrl 67/33 films which has an identical FA/Cs ratio and similar Cs loading to the DMA10 thin films; however, in these films we do not observe the same level of local FA heterogeneity as in the DMA10 samples (Figure S23). Together, these results indicate that the addition of DMA via this deposition method results in increasing local chemical compositional and structural heterogeneity, as reflected by increased PL linewidth, increased local spatial variations in the PL spectra, increased *d*-spacing heterogeneity and increased variation in the FA composition. This heterogeneity is observed both laterally and vertically, with inhomogeneous lateral distribution and the observation of increased DMA vs FA/Cs at the film surface. Taking these observations together, we propose that while DMA is incorporated throughout the perovskite material (as confirmed both by TOF-SIMS and the bulk increase in bandgap), when using this solution-based deposition method, perovskite material with even richer DMA content nevertheless forms at the surface, in an inhomogeneous manner.

The conclusion that we observe both vertical and lateral inhomogeneity in DMA-incorporated perovskite films is unexpected. Films made in the same way and of the same composition show enhanced device performance.[24,25] It is worth noting that we carried out these measurements on films fabricated both at National Renewable Energy Lab (where devices with improved performance are fabricated) and at University of Washington, with effectively identical results on films made in both places. Typically, we expect less homogeneous materials to result in less well-performing devices.[15,16] However, similar observations of heterogenous incorporation simultaneous with performance improvement have also been reported by Fenning and co-workers for smaller alkali metals, such as Rb, incorporation at the A-site, so these observations are not unprecedented.[42] Taking our observations along with the observations of enhanced device performance we have made previously, would suggest that the higher efficiencies and greater stabilities of the DMA-incorporated films arise at least in part, due to surface passivation from the DMA. However, the heterogeneity of the DMA incorporation laterally is likely suboptimal for a device performance standpoint, reducing the performance from what it could be with perfectly homogeneous passivation, and causing voltage loss due to the broadening of the PL. So while the benefits of the DMA incorporation do outweigh the downsides in terms of device performance,[24,25] our work here suggests that if the DMA could be more homogeneously incorporated, further performance gains would be expected. Our observations are promising news for the use of DMA as an additive, as they indicate that there is room for improving the current processing conditions that leads to more homogeneous chemical composition in the DMA incorporated films. More homogeneous films incorporating DMA could in turn lead to further improvements in the device performance and operational stability. There is certainly scope for exploring both different ways of solution-processing the DMA-containing material, or more intrinsically homogenous processes such as vapor deposition could be very promising.

**Acknowledgements**

This material is based on the work primarily supported by the U.S. Department of Energy's Office of Energy Efficiency and Renewable Energy (EERE) under the Solar Energy Technology Office (SETO), Award Number DE-EE0008747. Part of this work was conducted at the Molecular Analysis Facility, a National Nanotechnology Coordinated Infrastructure site at the University of Washington which is supported in part by the National Science Foundation (grant NNCI-1542101), the University of Washington, the Molecular Engineering & Sciences Institute, and the Clean Energy Institute. SJ thanks Dan Graham at Molecular Analysis Facility at the University of Washington for his help with TOF-SIMS data collection and analysis. SJ also thanks Jessica Kong, Shaun Gallagher and Dr. Lucas Flagg for experimental help with PiFM and PL mapping.

# Supplementary Information for

# Dimethylammonium additives alter both vertical and lateral composition in halide perovskite semiconductors

Sarthak Jariwala[1,2], Rishi Kumar[3], Giles E. Eperon[4,5], Yangwei Shi[1], David Fenning[3], David S. Ginger[1*].

1. Department of Chemistry, University of Washington, Seattle, WA 98195, USA
2. Department of Materials Science and Engineering, University of Washington, Seattle, WA 98195, USA
3. Department of Nanoengineering, University of California, San Diego, La Jolla, CA 92093, USA
4. National Renewable Energy Laboratory, Golden, CO 80401, USA
5. Swift Solar Inc., San Carlos, CA 94070, USA

*Corresponding author – dginger@uw.edu

**Experimental Procedures**

Materials and Methods

Perovskite precursor solutions were prepared and deposited according to method reported by Eperon *et al.* [1,2] FAI (Greatcell), DMAI (Greatcell), CsI (Sigma Aldrich), $PbI_2$ (TCI), $PbBr_2$ (Sigma) and $PbCl_2$ (TCI) were dissolved in anhydrous DMF & DMSO (ratio of 3:1 v/v) to produce a 1M solution of $FA_xCs_{1-x-y}DMA_yPb(I_{0.8}Br_{0.2})_3$. The small amount of Cl added is in stoichiometric excess (6 mg for a 1mL solution).

Perovskite solutions of varying A site composition (DMA10, Ctrl 67/33 and Ctrl 80/20) were made by tuning the amount of FAI, CsI and DMAI according to the formula above.

All precursors were used as purchased and stored and mixed in a Nitrogen glovebox.

Perovskite thin films were prepared on plasma/UV-Ozone cleaned glass. The solution was deposited on top of the substrate and spun at 5000 rpm for 60 s. At ~35 s remaining, anhydrous methyl acetate antisolvent was dropped from above. The thin films were then annealed at 100 C for 30 mins. The thin films were prepared in a Nitrogen filled glovebox.

Samples were prepared at National Renewable Energy Laboratory (NREL) as well as University of Washington (UW) and characterized at UW.

Time-of-flight secondary-ion mass spectrometry (TOF-SIMS)

TOF-SIMS data was collected using IONTOF TOF-SIMS 5. $Bi^{3+}$ was used as primary ion for TOF-SIMS analysis. The primary ion dosage used for depth profile was ~1.5 x $10^{11}$ ions/$cm^{-3}$. The measurement was performed in high mass resolution mode. The sputtering was done using Ar (at 20 keV) with a cluster size of 1000. The sputtering ion dosage used was ~6.1 x $10^{13}$ ions/$cm^{-3}$. An area of 500 x 500 um was sputtered and a raster size of 100 x 100 um within the sputtered area was used for analysis. Identical settings were used for both positive and negative ions.

Photoluminescence Microscopy

Local photoluminescence (PL) spectra were collected using a custom scanning confocal microscope built around a Nikon TE-2000 inverted microscope. An infinity corrected 100× dry objective (Nikon LU Plan Fluor, NA 0.9) was used for the acquisition. The sample was illuminated with a 470 nm pulsed diode laser (PDL-800 LDH-P-C-470B, 2.5 MHz, ~300 ps pulse width) and the sample PL was filtered using a 600 nm long pass filter and directed onto a Ocean Optics spectrometer (USB-2000). The sample stage was controlled using a piezo controller (Physik Instrumente E-710) and the pixel dwell time was 10-100 ms.

Synchrotron X-Ray Diffraction and Fluorescence

Nano-XRF and nano-XRD were both acquired at the 26-I*D*-C beamline at the Advanced Photon Source at Argonne National Laboratory using a focused 9 keV X-ray beam. *d*-spacing was determined precisely by performing a spatially-resolved rocking curve, where the incident and detector angles were held constant in the Bragg condition while the sample angle with respect to the beam was rotated ($\omega$ scan). The 2D scattering detector was calibrated by measurement of a powder silicon reference sample. This procedure enables complete reconstruction of the 3D scattering vector at each point, the magnitude of which is proportional to the *d*-spacing. XRF measurements were corrected for attenuation of both the incident and fluoresced X-rays, using the nominal stoichiometry ($Cs_{0.3}FA_{0.6}DMA_{0.1}PbI_{2.4}Br_{0.6}$ for DMA, $Cs_{0.2}FA_{0.8}PbI_{2.4}Br_{0.6}$ for Control 80/20) and a density of 4.42 g cm$^{-3}$ to approximate the X-ray attenuation coefficient of the sample. The fluorescence detection was calibrated using National Institute of Standards and Technologies (NIST) XRF standards (NBS1832, NBS1833).

Extracting Grain Size from Nanodiffraction Data

The grain size is determined by segmentation and measurement of the spatial scattering intensity at each diffracting point in reciprocal space. Before segmentation, however, the projection of the incident X-ray probe is deconvoluted from the spatial scattering intensity. The diffraction data, acquired with the sample in the Bragg condition, requires the incident beam to be at a shallow angle (~ 14°) relative to the sample surface, causing a horizontal "blur" of about two microns in our sample and measurement geometry. This "blur" is defined as a point-spread function (PSF) weighted by the relative signal intensity from each point in the sample, and is quantified by calculating the attenuation of the 9 keV x-ray probe along the path length into the sample and out to the detector. The influence of this PSF is removed from the spatial scattering intensity by Richardson-Lucy deconvolution[3] to yield true maps of diffracting intensity (Figure REKSI10). Finally, these diffracting intensity maps are segmented into distinct regions, whose areas are tabulated. This process is repeated for each diffracting region of reciprocal space. In this way, diffracting regions are identified that are separated in both real and reciprocal space and therefore can be considered an individual diffracting grain.

Photo-induced Force Microscopy (PiFM)

A Molecular Vista VistaScope coupled to a LaserTune QCL was used for PiFM imaging. The PiFM was operated in sideband mode where the first and second mechanical resonance frequencies at 320 kHz and 1.9 MHz of a Si cantilever with Pt coating were used to detect the photoinduced force and surface topography, respectively. The spatial resolution of the hyperspectral image acquired was 256 × 256 pixels /512 x 512 pixels. The PiFM spectra was collected by sweeping the laser from 760 to 1875 cm$^{-1}$. The PiFM imaging was then carried out by keeping the laser at 1714 cm$^{-1}$, corresponding to the C=N stretch in FA (formamidinium).

Data Analysis

TOF-SIMS data were analyzed using commercial software provided by IONTOF. PL microscopy data were analyzed using an inhouse software that is openly available online at https://github.com/SarthakJariwala/Python_GUI_apps. All the data were then plotted and analyzed further in python using numpy,[4] matplotlib,[5] seaborn,[6] and seaborn-image (https://github.com/SarthakJariwala/seaborn-image).

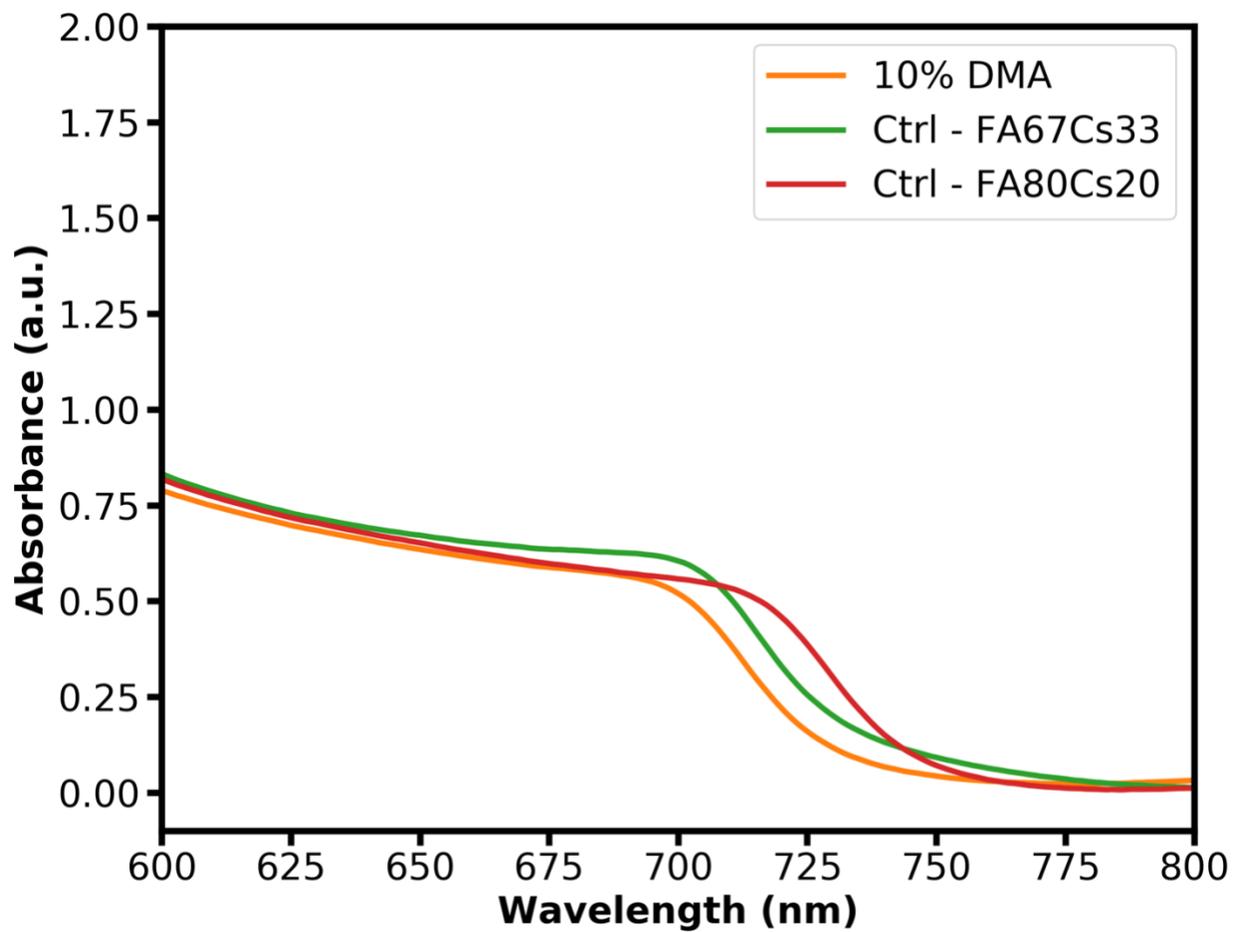

Figure S1. UV-Vis absorption spectra of DMA10, Ctrl 67/33 and Ctrl 80/20 thin films.

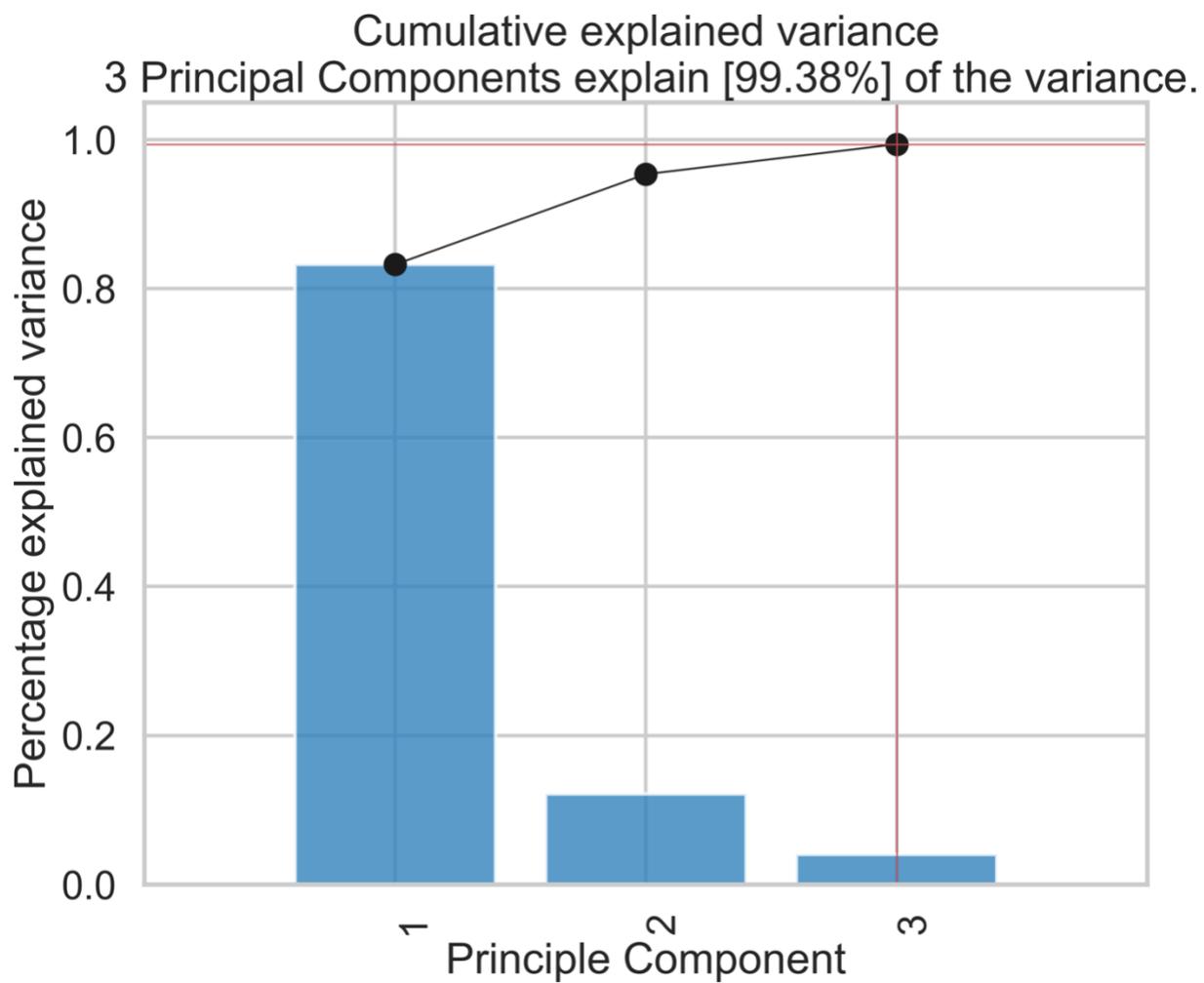

Figure S2. Cumulative explained variance observed in the TOF-SIMS spectra of DMA 10, Ctrl 67/33 and Ctrl 80/20 thin films. 3 principal components explain over 99% of the observed variance in the TOF-SIMS spectra.

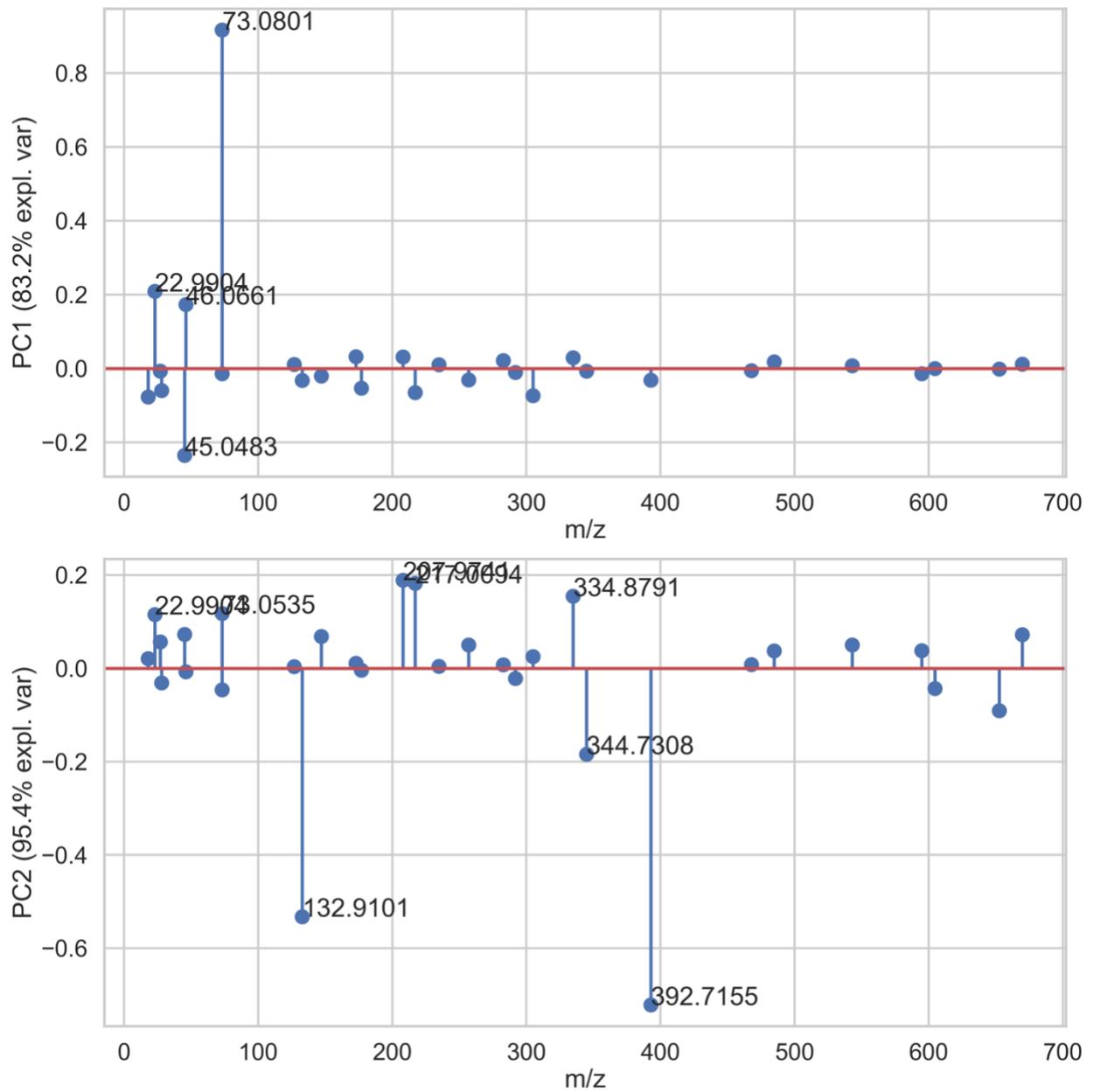

Figure S3. Plot of loadings for the scores of first (top) and second (bottom) principal component. Only peaks contributing to >0.1 are labelled in the figure.

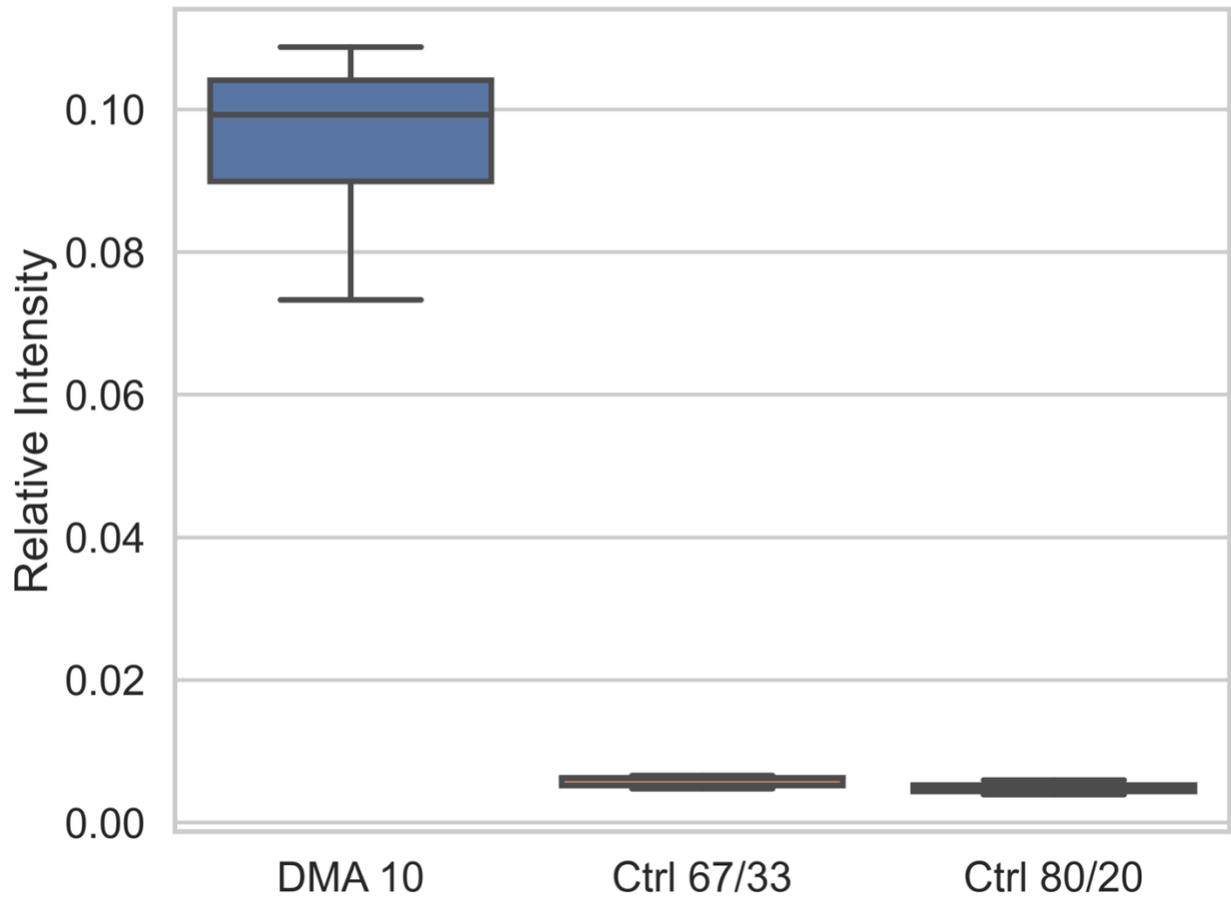

Figure S4. Normalized relative intensity of C3H9N2+ peak spectra (m/z = 73.08) for DMA10, Ctrl 67/33 and Ctrl 80/20 thin films.

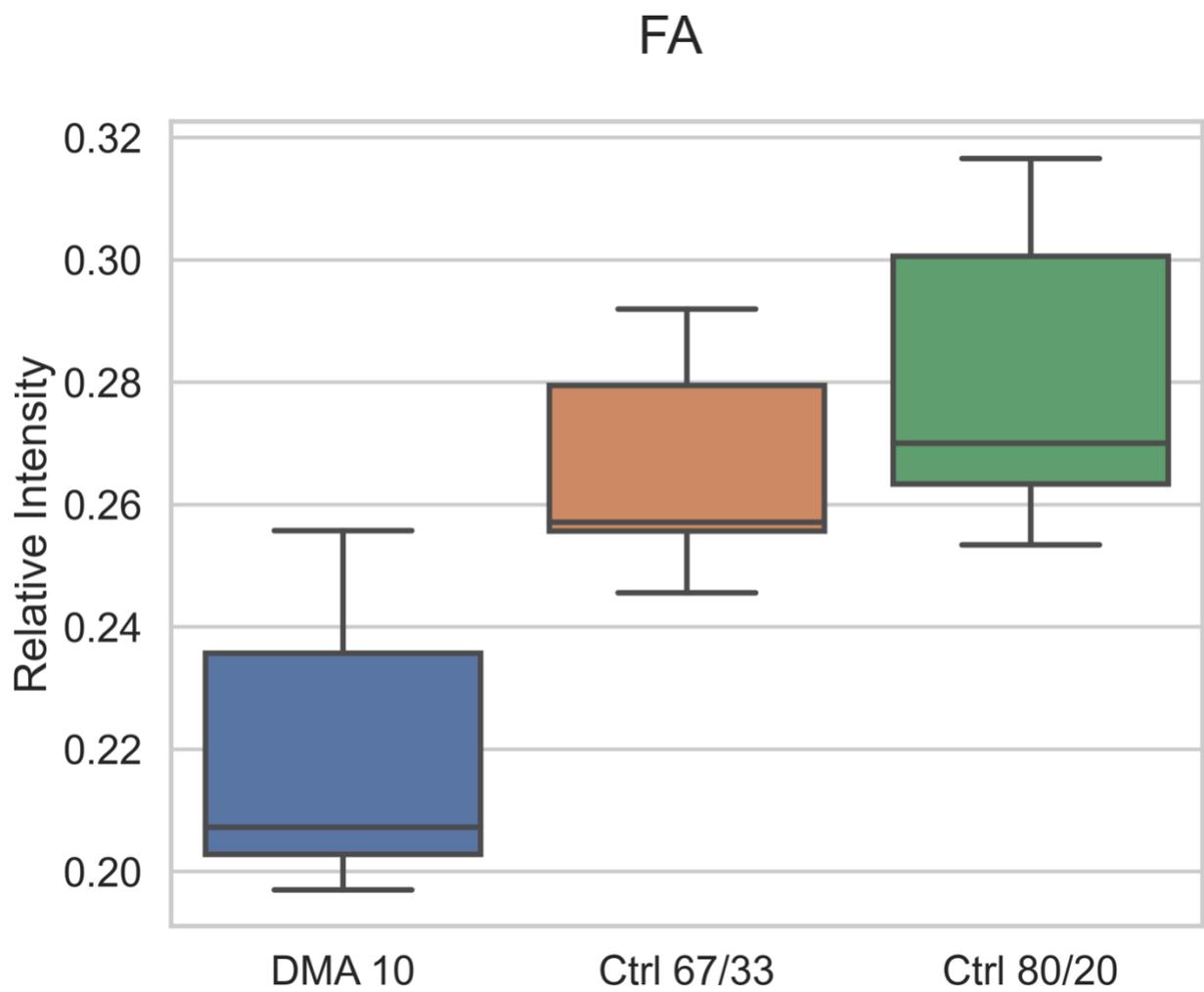

Figure S5. Normalized relative intensity of FA peak spectra (m/z = 45.05) for DMA10, Ctrl 67/33 and Ctrl 80/20 thin films.

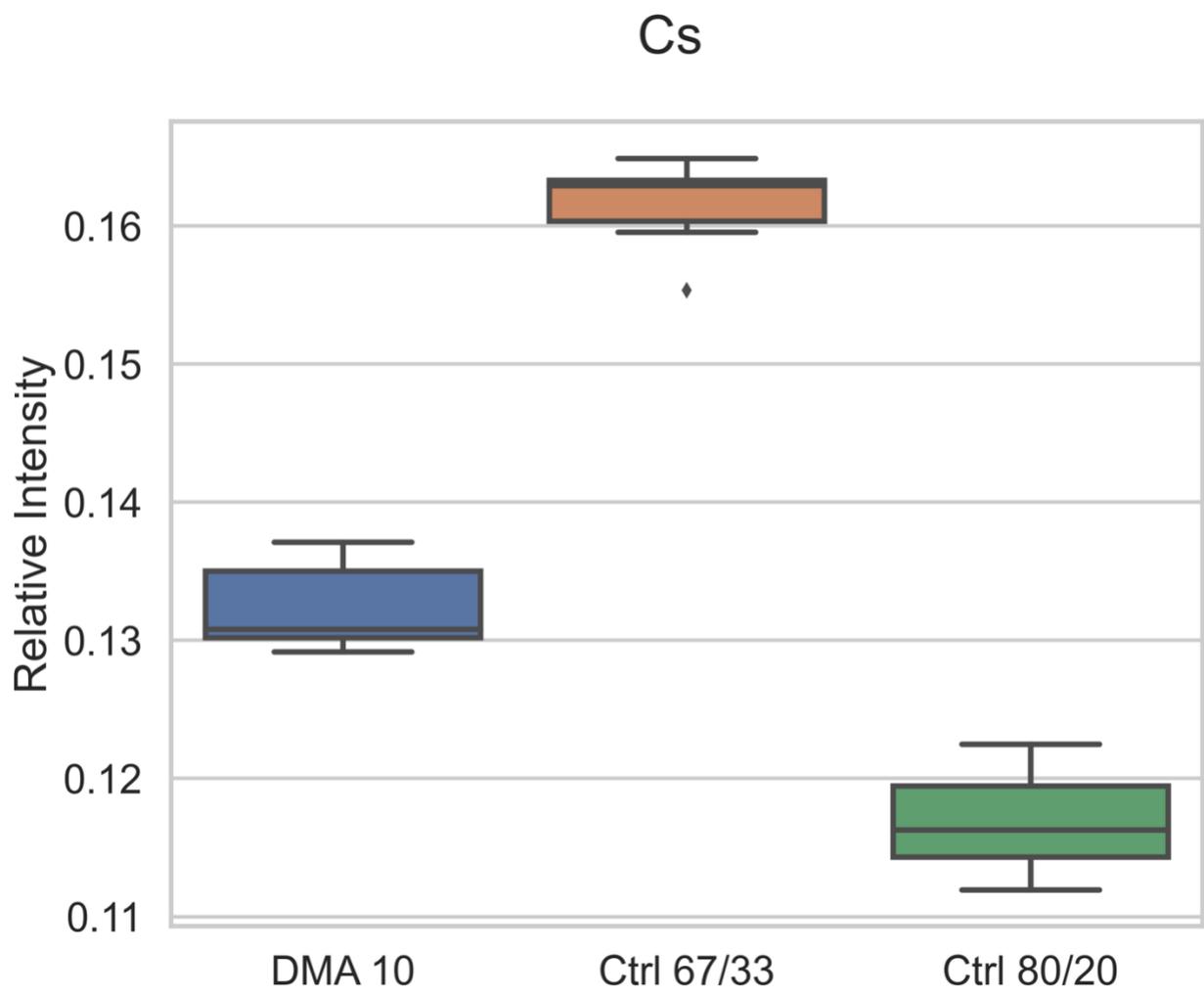

Figure S6. Normalized relative intensity of Cs peak spectra (m/z = 132) for DMA10, Ctrl 67/33 and Ctrl 80/20 thin films.

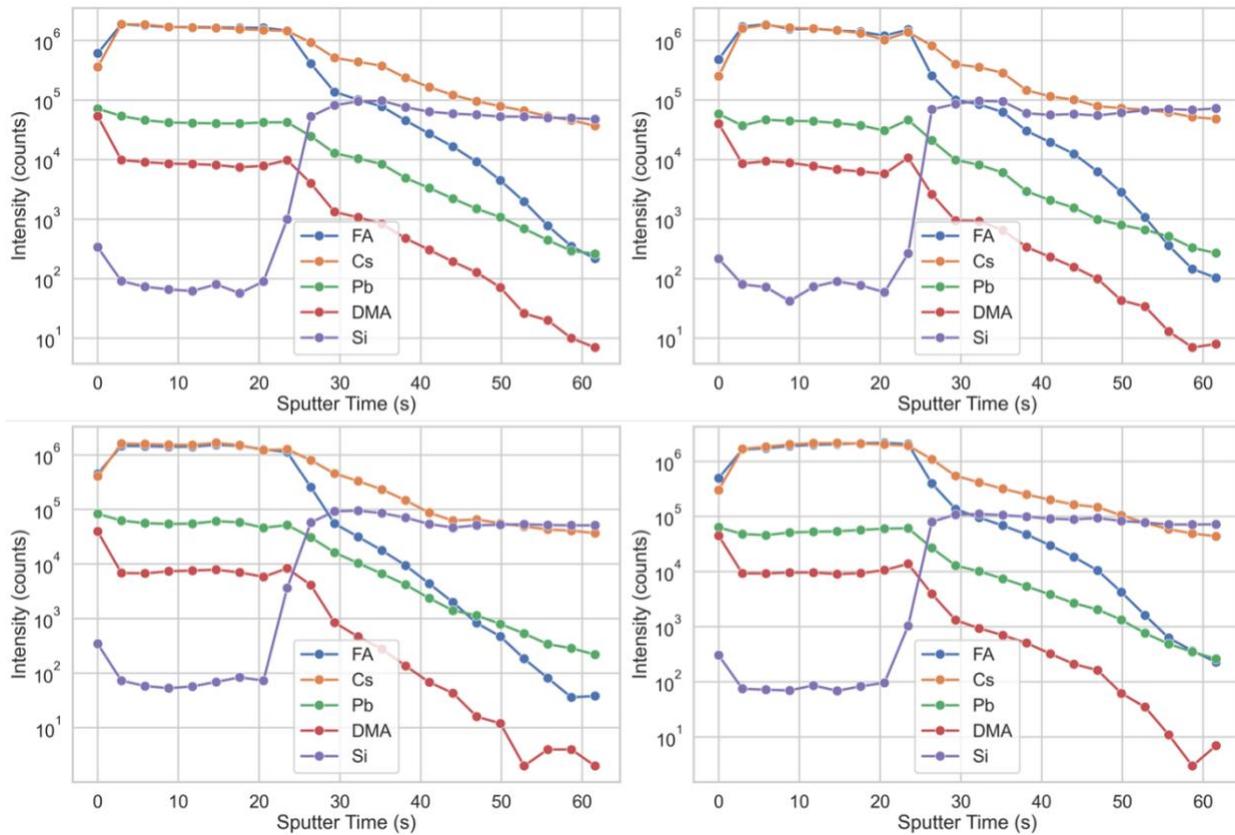

Figure S7. TOF-SIMS depth profile of positive ions – FA, Cs, Pb, DMA, and Si acquired on different regions of different DMA10 samples.

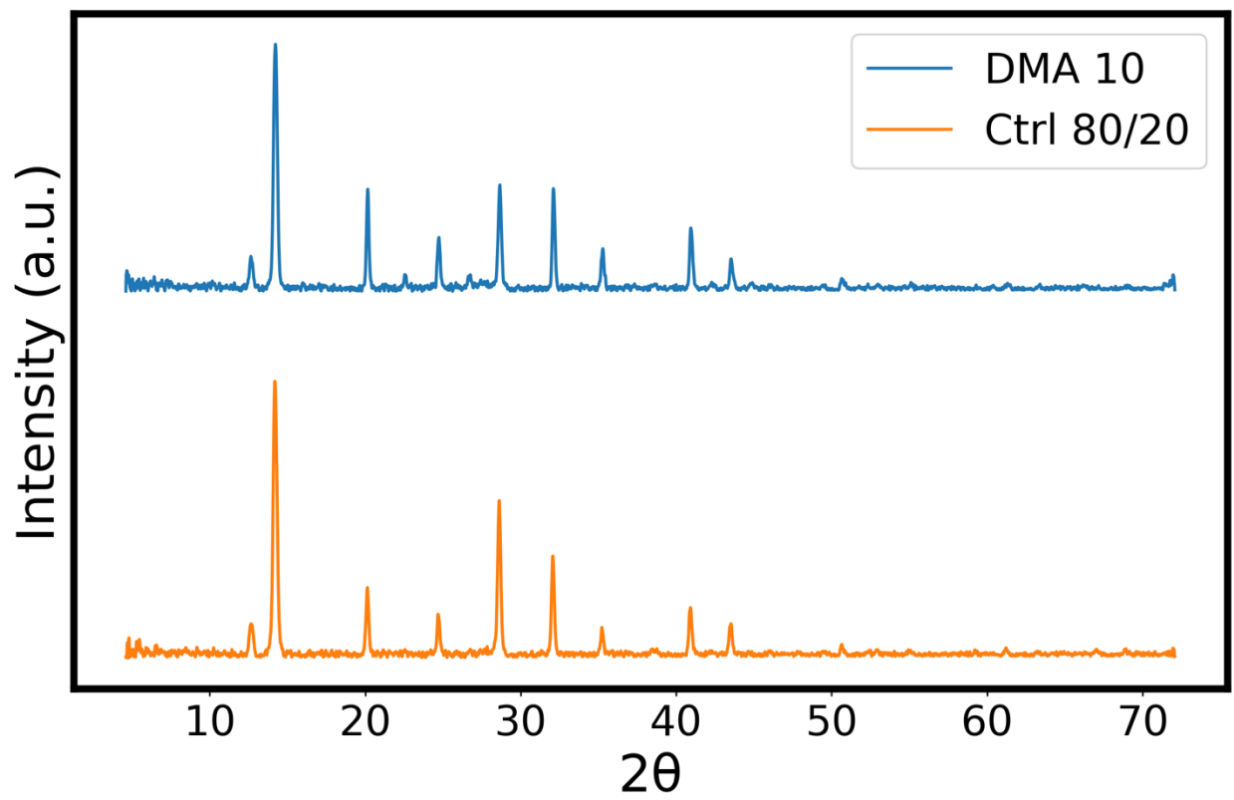

Figure S8. XRD of DMA10 thin films.

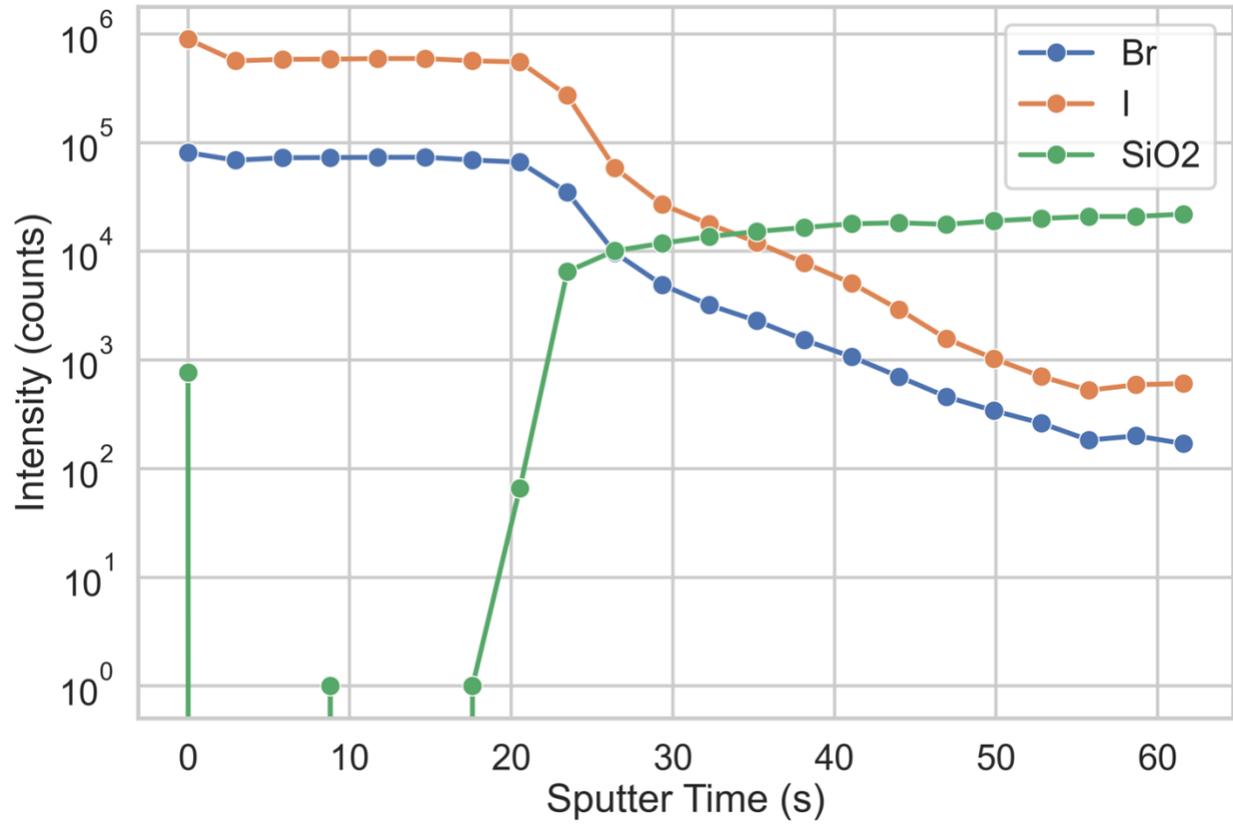

Figure S9. TOF-SIMS depth profile of negative ions – Br, I and SiO$_2$ acquired on a DMA10 thin film.

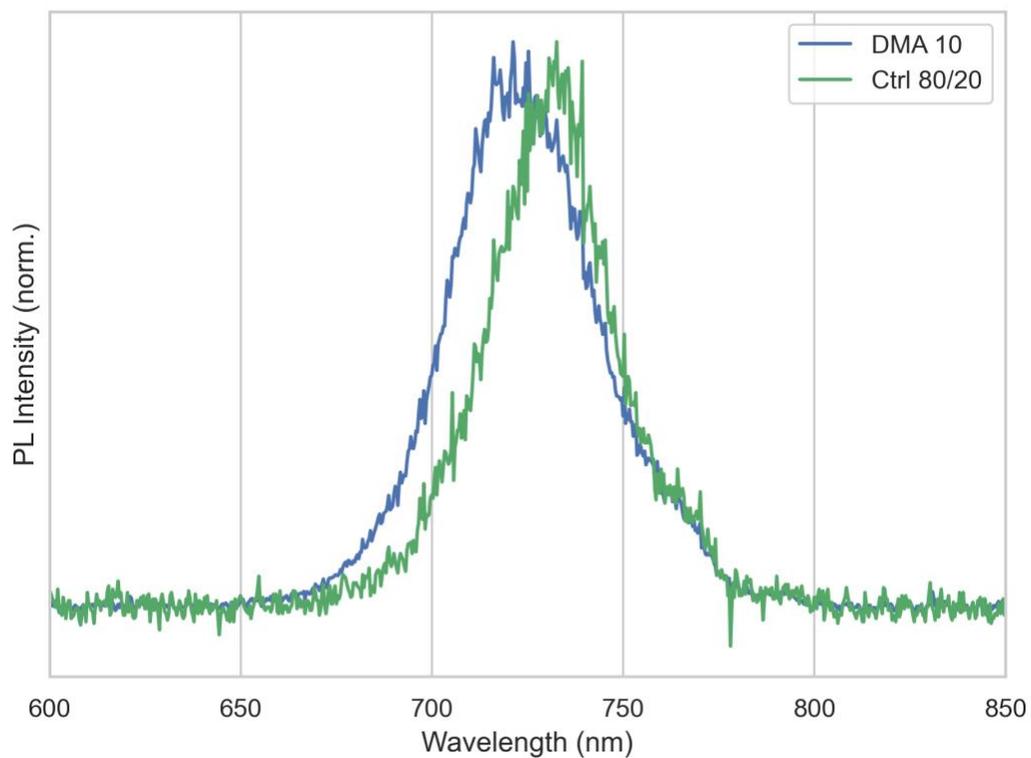

Figure S10. Bulk PL spectra collected on representative DMA 10 and Ctrl 80/20 thin films showing a higher FWHM for DMA 10 compared to Ctrl 80/20. See main text for more discussion on the underlying origin of the increased FWHM.

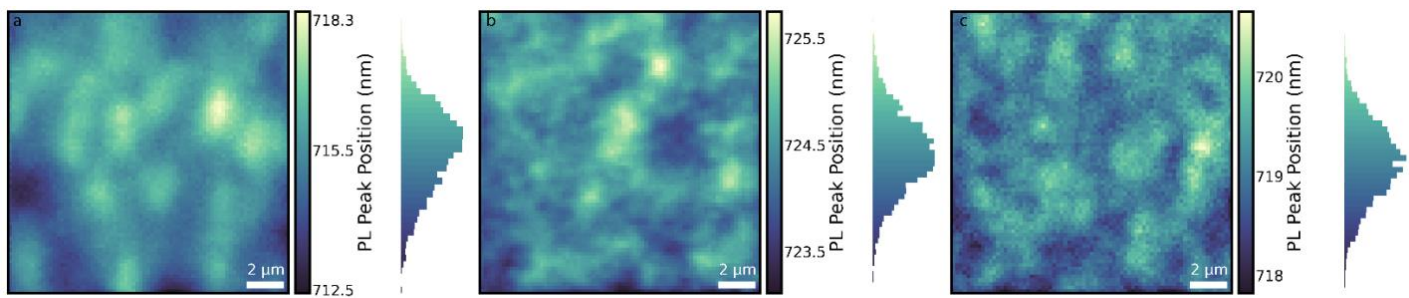

Figure S11. PL spectroscopy images of (a) DMA10, (b) Ctrl 80/20, and (c) Ctrl 67/33 thin films, local PL spectra were acquired at each pixel. See Figure 2 for normalized images.

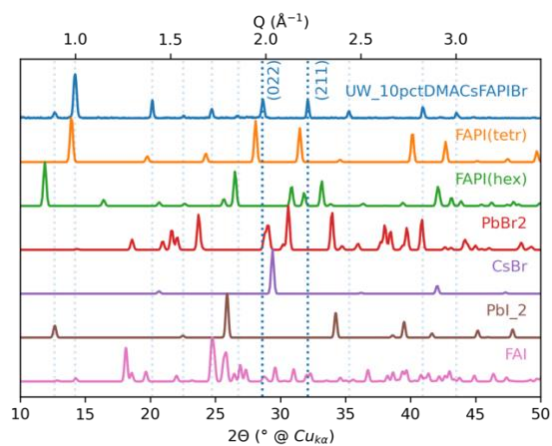

Figure S12. Benchtop XRD patterns for the DMA containing sample and related phases that could be present due to either phase segregation or perovskite decomposition. Dashed lines indicate peaks corresponding to the target sample. The two darker lines indicate the (022) and (211) peaks, which were analyzed using synchrotron nanoXRD. These peaks do not overlap with any peaks from precursors or segregated phases and can be attributed specifically to our target phase.

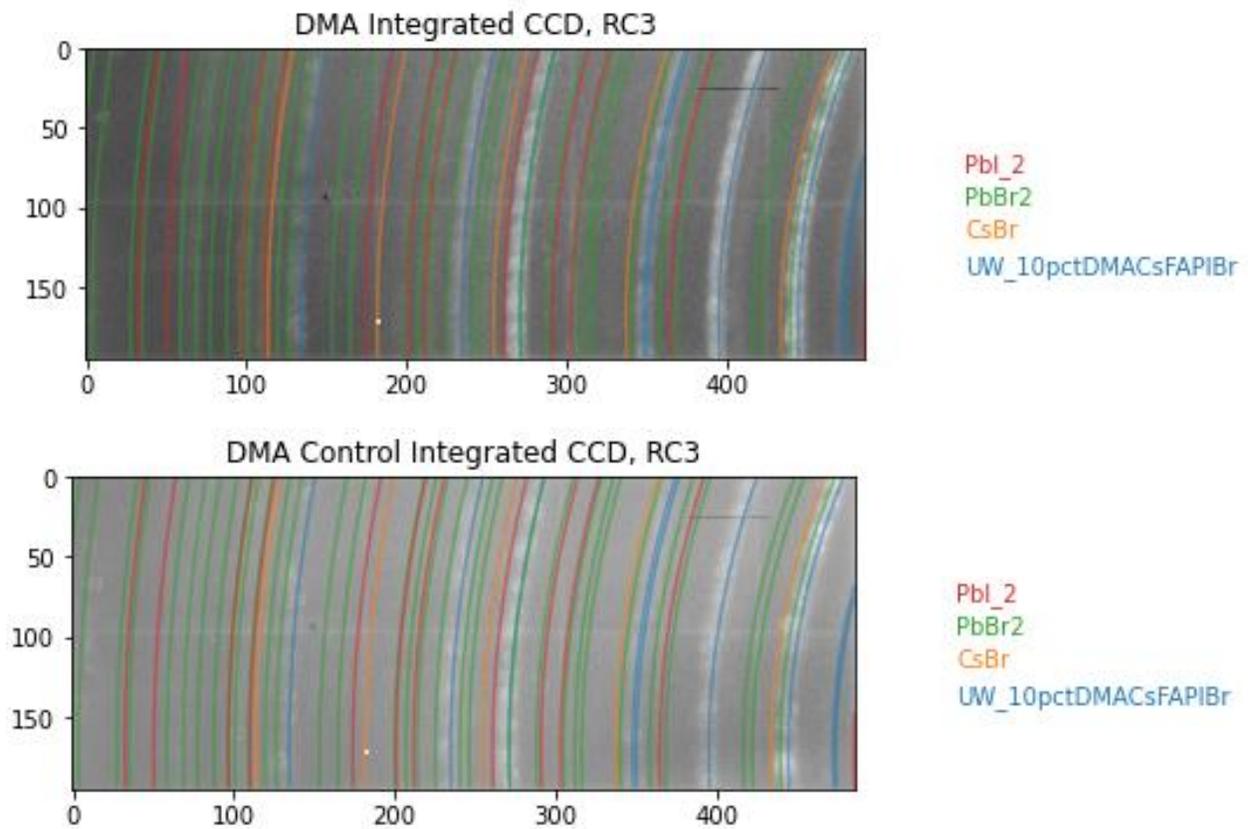

Figure S13. 2D panel detectors (showing reciprocal space) integrated over a 30x30 µm region of DMA (top) and Control 67/33 (bottom) samples. Rings demarcate angles expected for possible phases based on benchtop XRD.

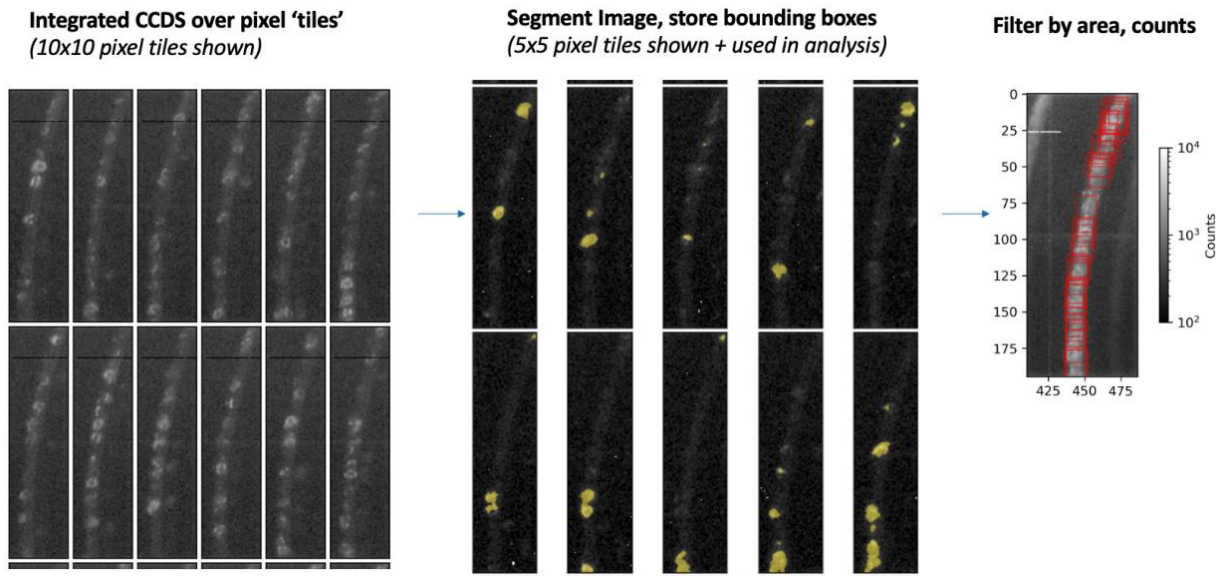

Figure S13. Isolation of diffracting grains by segmentation in both real and reciprocal space. The diffraction intensity is integrated over small (2.5 x 2.5 μm) regions of realspace (first, second figures). In these reciprocal space images, each diffraction "spot" represents a single diffracting grain (yellow overlay in center figure). By segmenting these images, regions of interest are established in reciprocal space (red boxes in third figure).

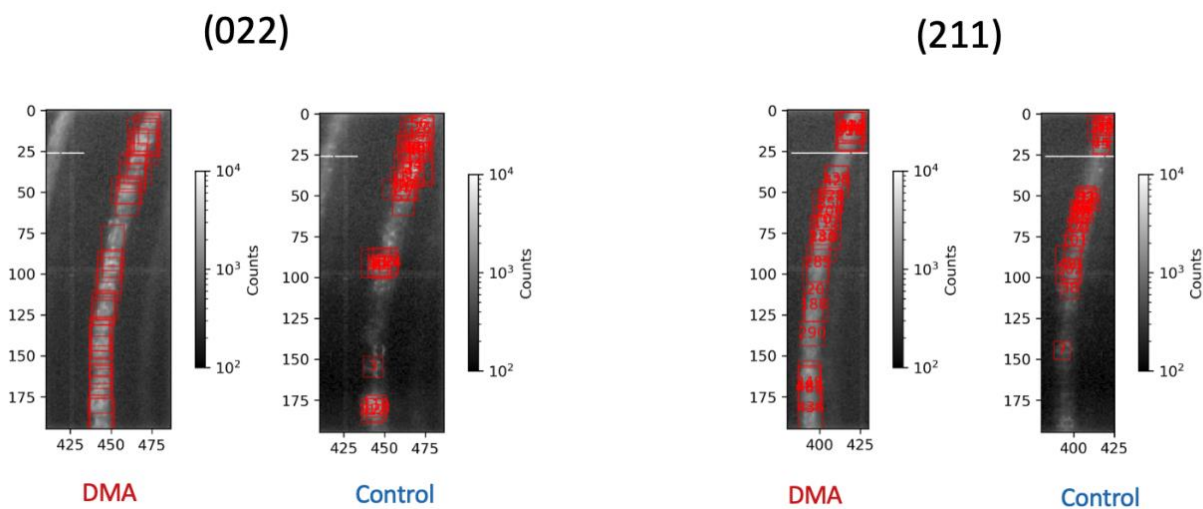

Figure S14. Red boxes are placed around regions of interest (ROIs) of reciprocal space for which *d*-spacing and grain size are analyzed.

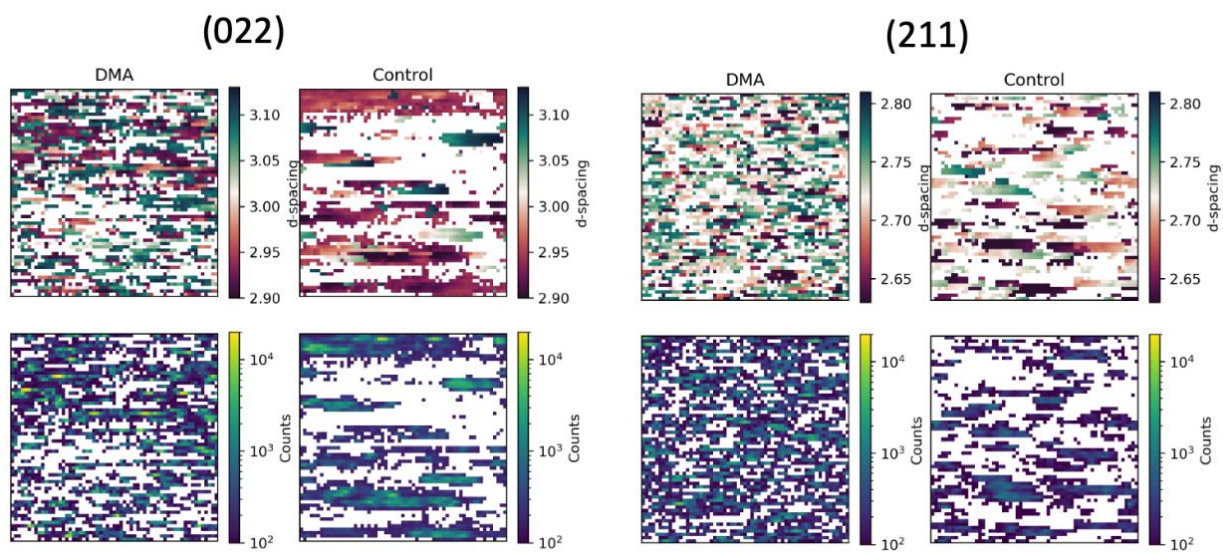

Figure S15. Real-space maps of *d*-spacing (top row) and diffraction intensity (bottom row) for DMA and Control 67/33 samples. These maps are composed of maps at each of the reciprocal space ROIs in Fig. REKSI3. Regions with insignificant diffraction intensity are excluded and shown as white.

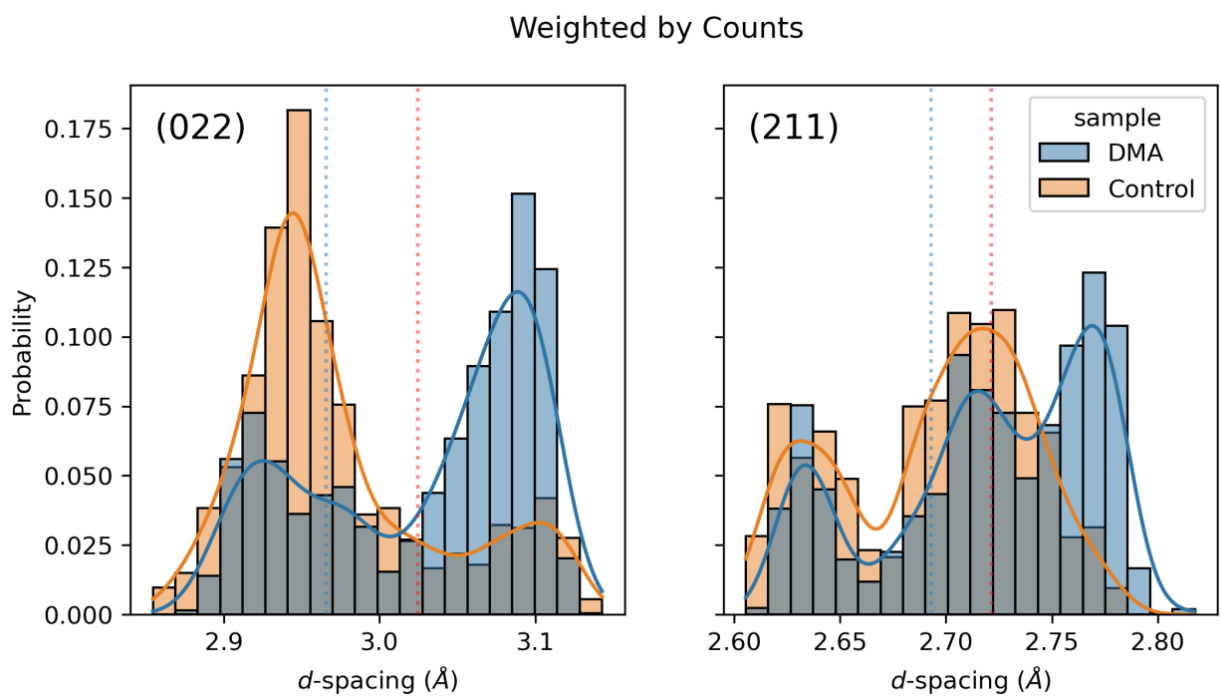

Figure S16. Distributions of *d*-spacing across the (022) and (211) reflections (same maps as Fig. REKSI4). Vertical lines indicate the mean *d*-spacing of each distribution.

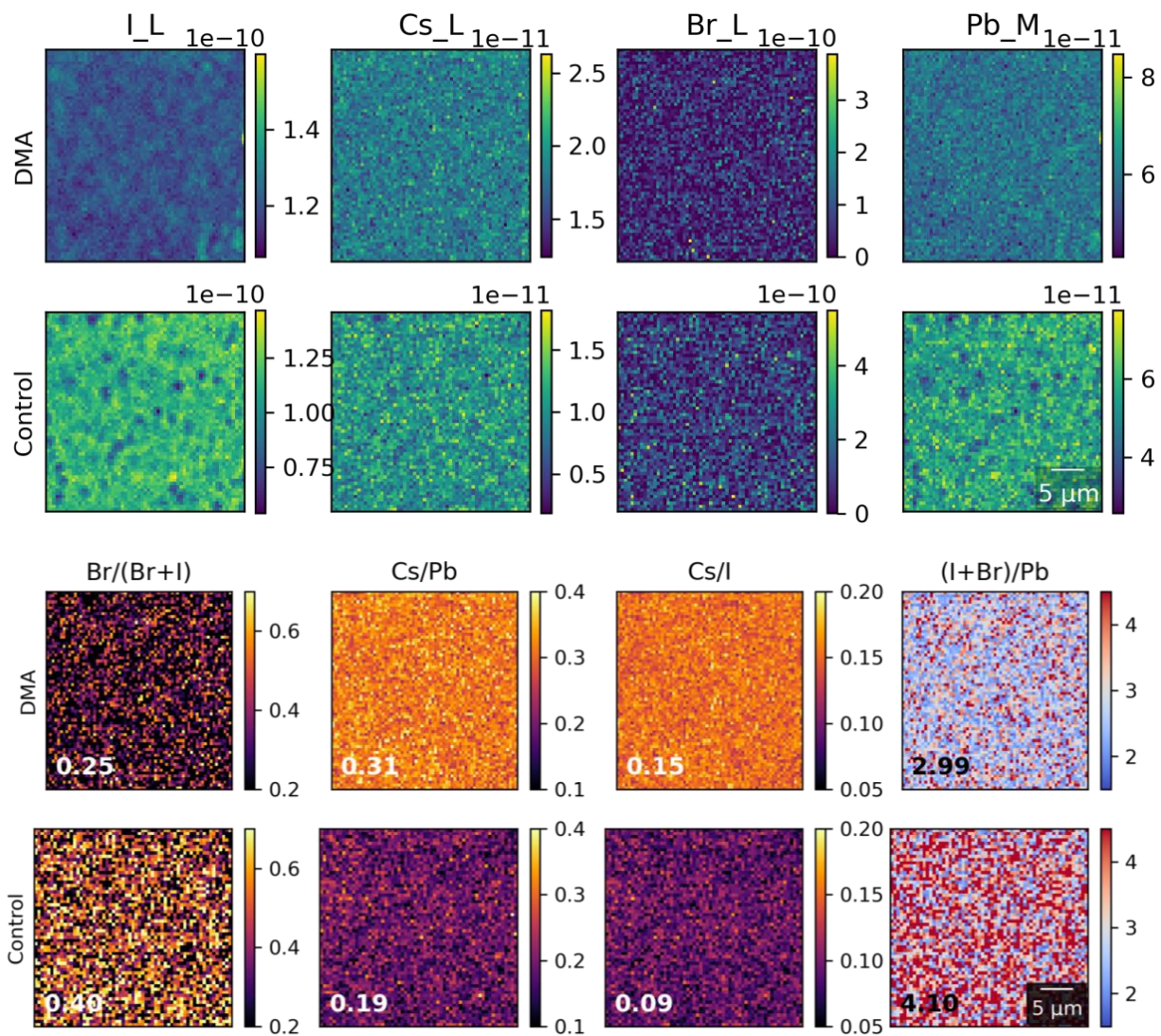

Figure S17. Areal density and ratios of elements present in the films. These are the same regions analyzed for *d*-spacing and grain size. The top two rows give individual elemental concentration in mol/cm². The bottom two rows give the molar ratios indicated by the above labels, and the average ratio across the map is indicated in the bottom left corner.

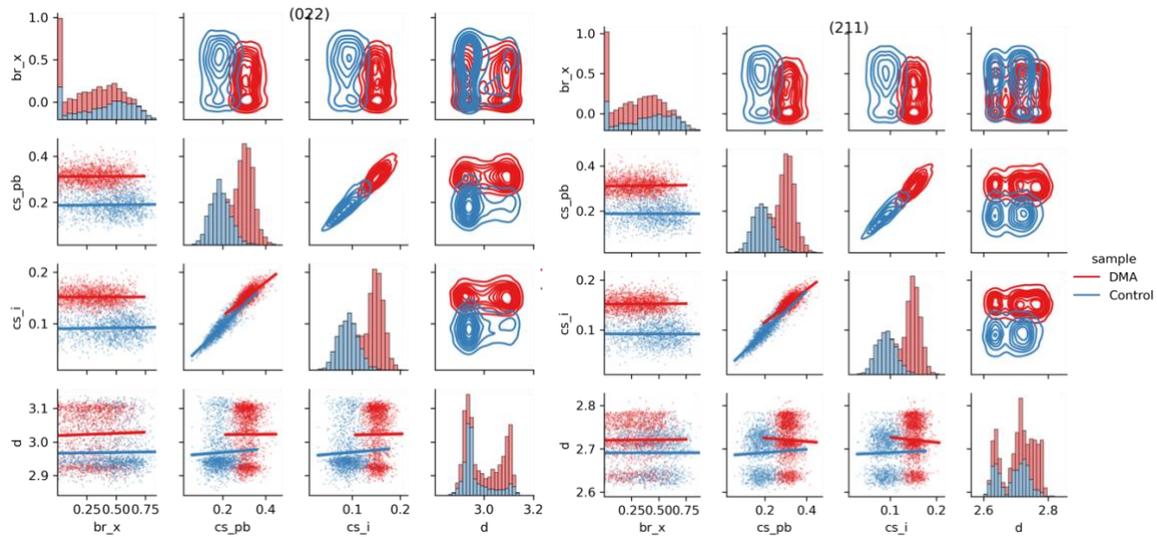

Figure S18. Spatial correlation of elemental molar ratios (br_x = Br/(Br+I), cs_pb = Cs/Pb, cs_i = Cs_I) and *d*-spacing (angstroms). The lower-left triangle shows scatter plots with linear regressions, the upper-right triangle shows kernel density estimates of the same data, and the diagonal shows histograms of the individual channel distributions.

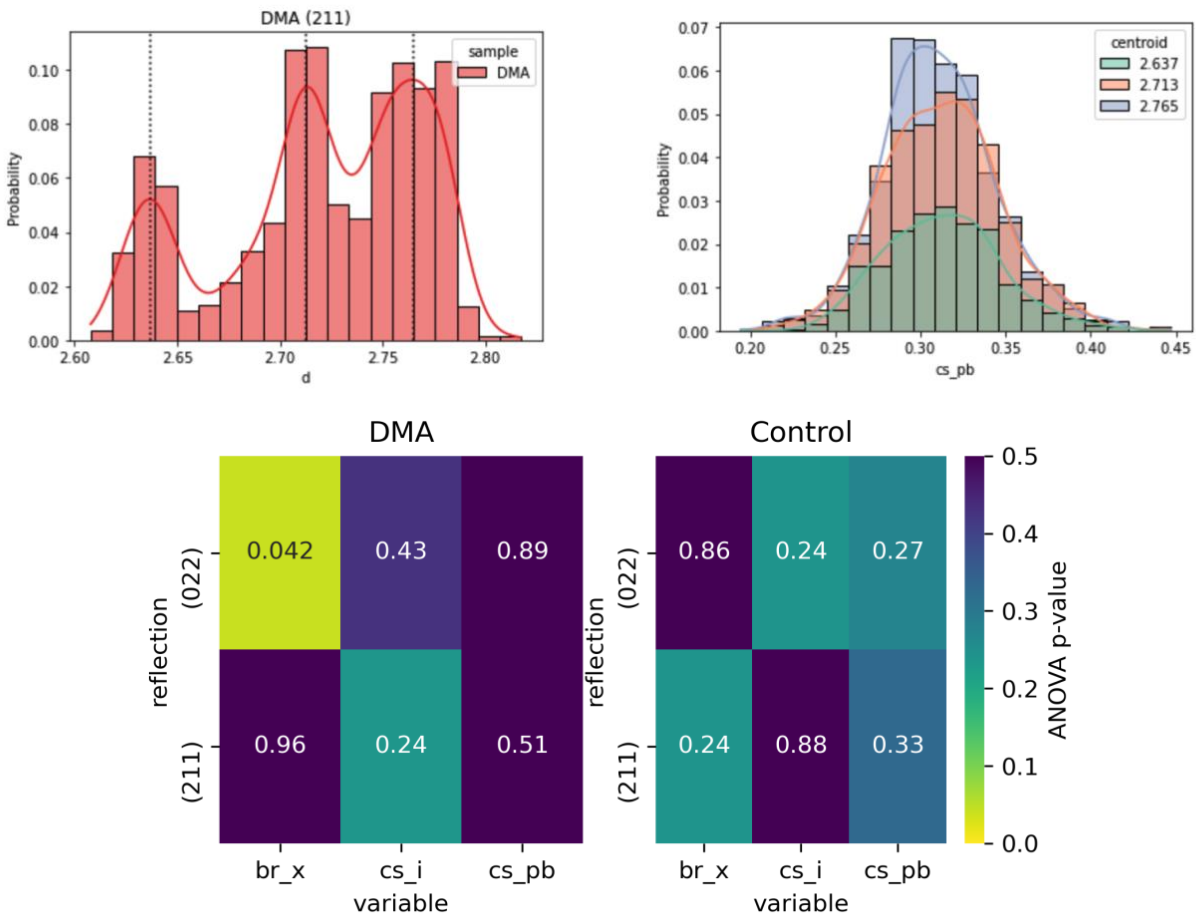

Figure S19. Each pixel of the nXRD + nXRF map is classified into one of the *d*-spacing distribution modes by proximity to the distribution means (a). This yields two (in control) or three (in DMA) populations of points classified by nXRD (b). The nXRF distributions of elemental ratios are compared across these populations by Analysis of Variance (ANOVA) tests to determine whether the composition varies significantly between *d*-spacings. The p-values resulting from this ANOVA test against various elemental ratios are summarized in (c).

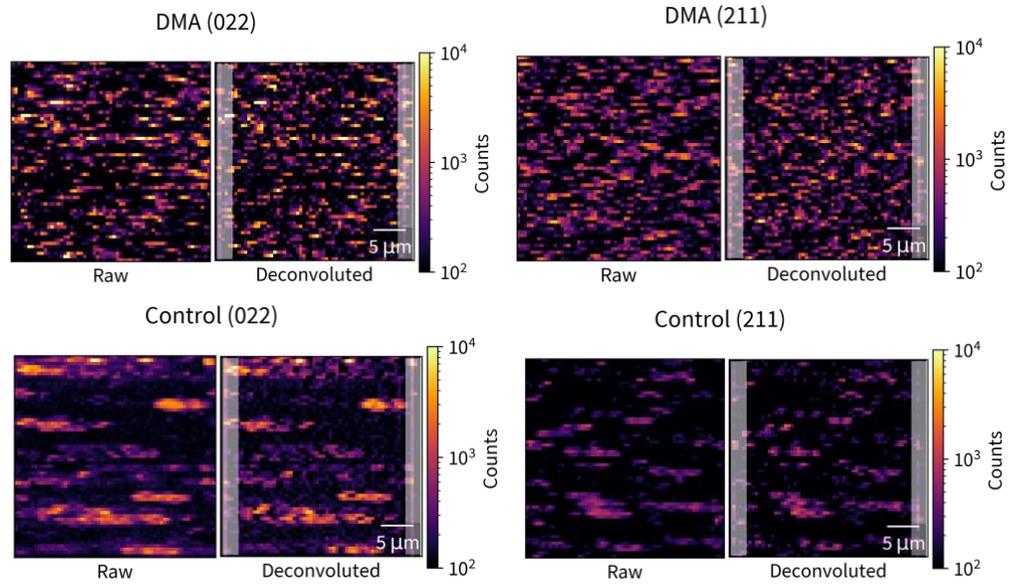

Figure S20. Spatial diffraction intensity integrated over all reciprocal space ROIs for (022) and (211) reflections in DMA and Control 80/20 samples. The raw counts are shown at left, while the probe-deconvoluted counts are shown at right. The white bands in the deconvoluted images can be affected by edge artifacts during the deconvolution process, and are excluded from subsequent analyses of grain size.

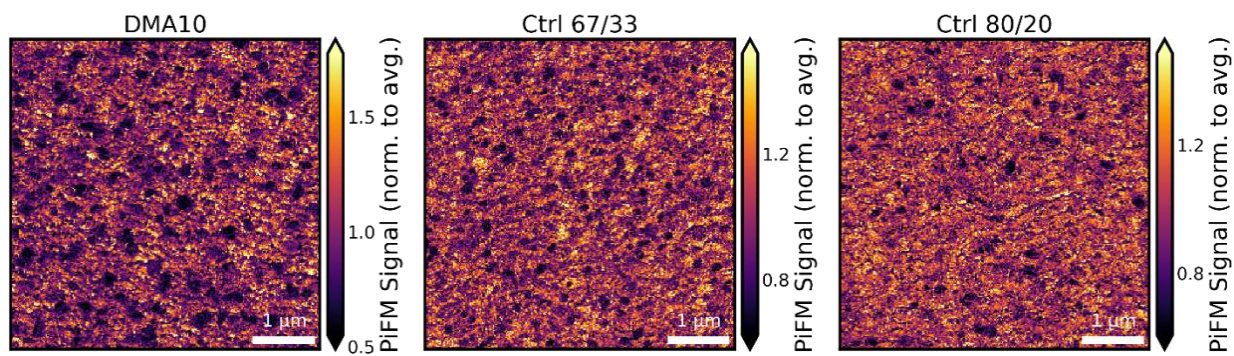

Figure S21. PiFM scan at 1714 cm$^{-1}$ for different DMA10, Ctrl 67/33 and Ctrl 80/20 thin films normalized to the average intensity showing larger variations in FA composition (1714 cm$^{-1}$) for DMA incorporated thin films.

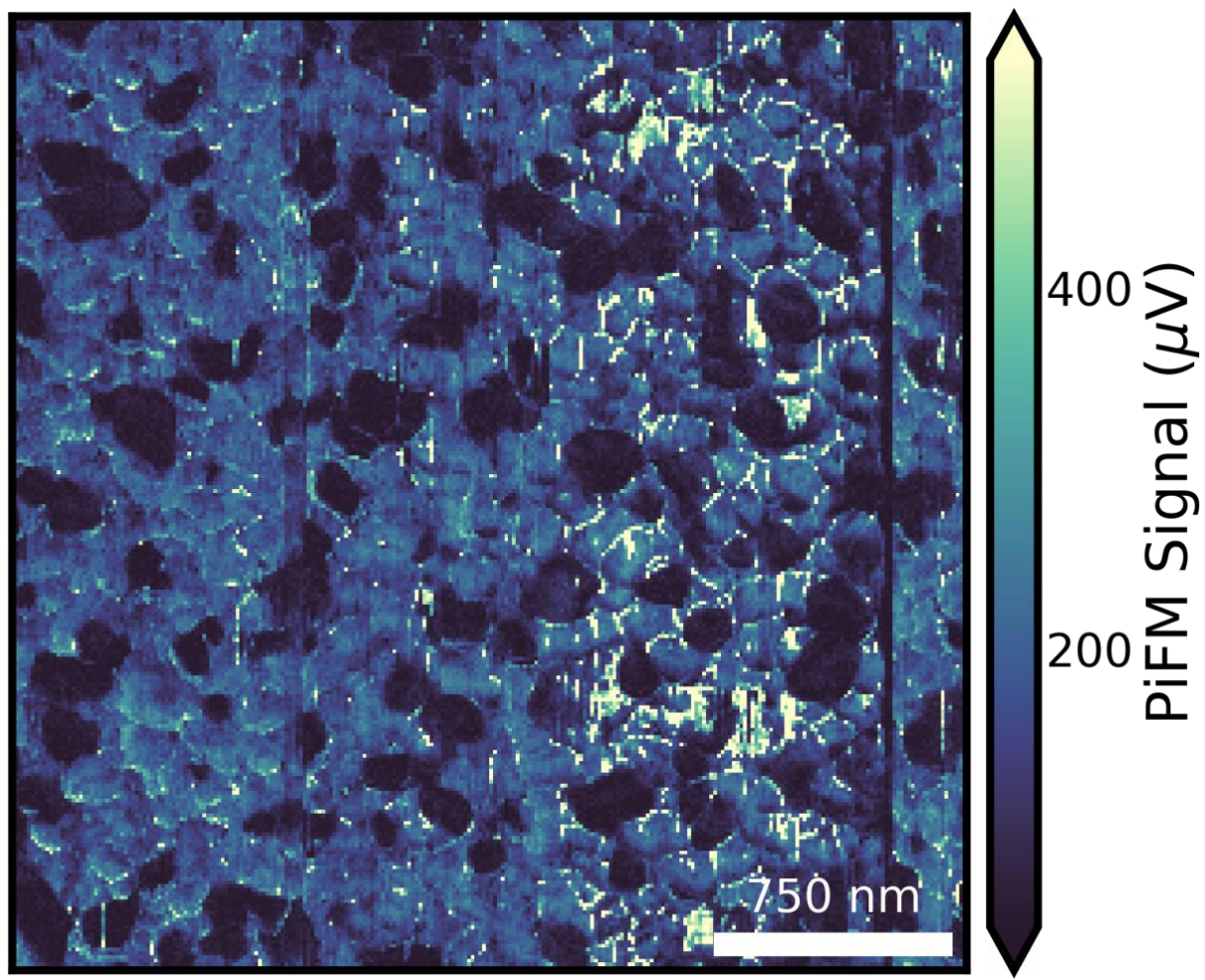

Figure S22. Higher spatial resolution PiFM image showing heterogeneity in the local FA composition in DMA10 film, acquired at 1714 cm$^{-1}$ (tracking local FA composition).

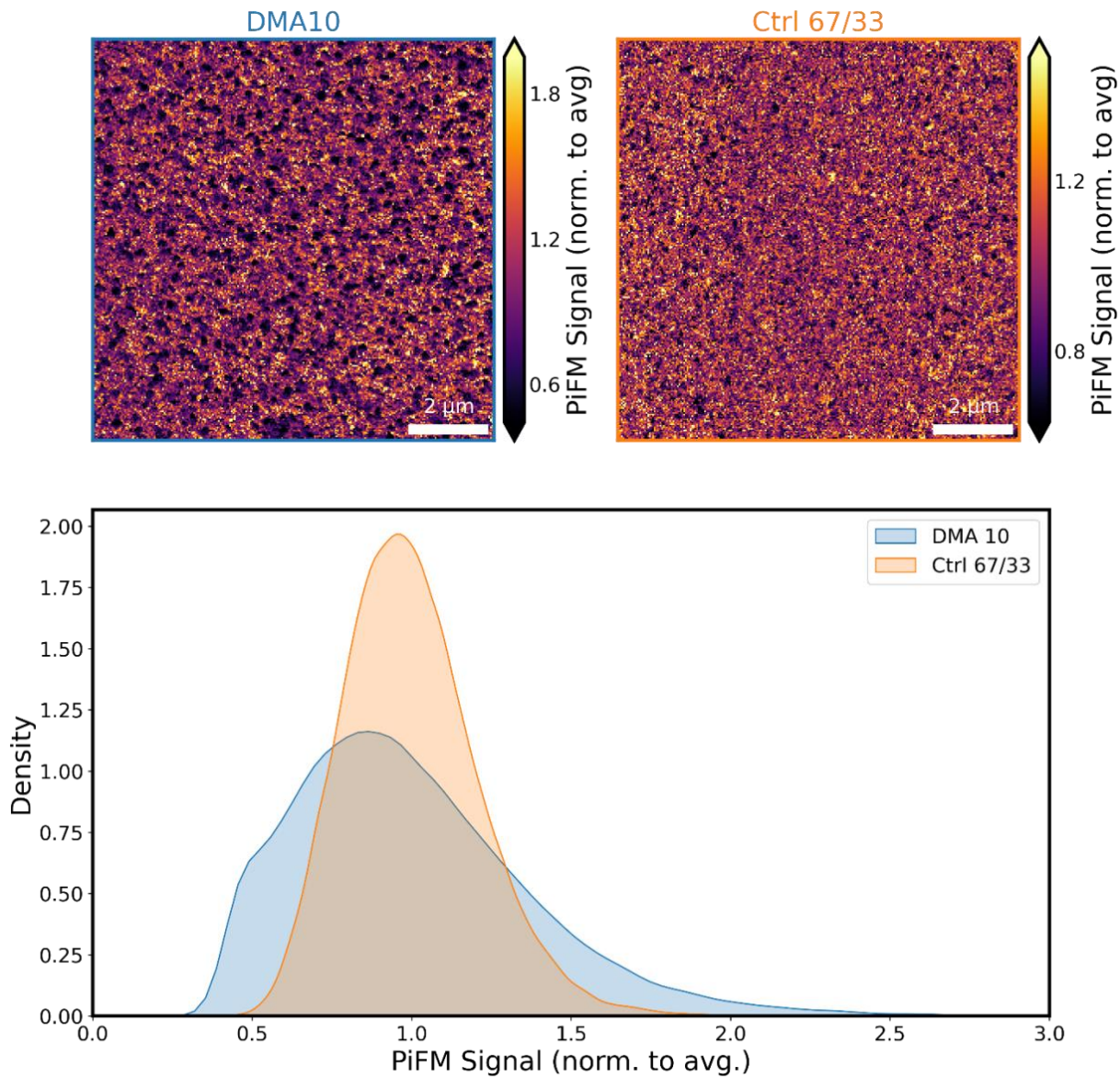

Figure S23. PiFM image at 1714 cm$^{-1}$ for various DMA10 (top left) and Ctrl 67/33 (top right) thin films showing that the distribution of FA intensity is more uniform in the Ctrl 67/33 thin films compared to DMA10 thin films (bottom). This indicates that the local FA heterogeneity is not merely a result of the increase in Cs concentration for the DMA10 films.